\documentclass[11pt,a4paper]{article}

\pdfoutput=1
\usepackage{jheppub}
\usepackage{amsfonts,amsbsy,amsmath,amssymb}
\usepackage{bbm}
\usepackage{color}
\usepackage{graphicx}
\usepackage[colorinlistoftodos]{todonotes}
\usepackage{booktabs}
\usepackage{subcaption}
\usepackage{multirow}
\usepackage{multicol}
\usepackage{here}
\usepackage[section]{placeins}
\usepackage[normalem]{ulem}
\usepackage{todonotes}
\usepackage{longtable}

\usepackage{bm}             
\usepackage{physics}
\usepackage{bbold}          
\usepackage{slashed}        
\usepackage{mathtools}      

\usepackage{pgf}
\usepackage{pgfplots}
\usepgfplotslibrary{dateplot}

\newcommand{\be}{\begin{equation}}
\newcommand{\ee}{\end{equation}}
\newcommand{\ba}{\begin{align}}
\newcommand{\ea}{\end{align}}
\newcommand{\baa}{\begin{array}}
\newcommand{\eaa}{\end{array}}
\newcommand{\bea}{\begin{eqnarray}}
\newcommand{\eea}{\end{eqnarray}}

\newcommand{\barm}{\overline{m}}

\newcommand{\basispl}{
   \put(-.5,-.5){\line(1,0){1}}
   \put(.5,-.5){\line(0,1){1}}
   \put(.5,.5){\line(-1,0){1}}
   \put(-.5,.5){\line(0,-1){1}}
                         }

\newcommand{\basisar}{
   \put(0,-.5){\vector(1,0){0}}
   \put(.5,0){\vector(0,1){0}}
   \put(0,.5){\vector(-1,0){0}}
   \put(-.5,0){\vector(0,-1){0}}
                      }

\newcommand{\plaq}{\setlength{\unitlength}{.5cm}\raisebox{-.2cm}{
   \begin{picture}(1.2,1.2)(-.6,-.6)
   \basispl\basisar
   \put(-.5,-.5){\circle*{.2}}
   \put(-.55,-.55){\makebox(0,0)[tr]{\footnotesize $n$}}
   \put(-.55,0){\makebox(0,0)[r]{\footnotesize $\nu$}}
   \put(0,-.55){\makebox(0,0)[t]{\footnotesize $\mu$}}
   \end{picture}}}
\newcommand{\twoplaq}{\setlength{\unitlength}{1cm}\raisebox{-.5cm}{
   \begin{picture}(1.2,1.2)(-.6,-.6)
   \basispl
   \put(-.5,-.5){\circle*{.1}}
   \put(-.5,.5){\circle*{.1}}
   \put(.5,-.5){\circle*{.1}}
   \put(.5,.5){\circle*{.1}}
   \put(0,-.5){\circle*{.1}}
   \put(0,.5){\circle*{.1}}
   \put(.5,0){\circle*{.1}}
   \put(-.5,0){\circle*{.1}}
   \put(-.25,-.5){\vector(1,0){0}}
   \put(.25,-.5){\vector(1,0){0}}
   \put(.5,-.25){\vector(0,1){0}}
   \put(.5,.25){\vector(0,1){0}}
   \put(-.25,.5){\vector(-1,0){0}}
   \put(.25,.5){\vector(-1,0){0}}
   \put(-.5,-.25){\vector(0,-1){0}}
   \put(-.5,.25){\vector(0,-1){0}}
   \put(-.55,-.55){\makebox(0,0)[tr]{\footnotesize $n$}}
   \put(-.55,0){\makebox(0,0)[r]{\footnotesize $\nu$}}
   \put(0,-.55){\makebox(0,0)[t]{\footnotesize $\mu$}}
   \end{picture}}}

\title{SU($N$) fractional instantons and the Fibonacci sequence}

\author[a,b]{Jorge Dasilva Golán,}
\author[a]{Margarita Garc\'{i}a P\'erez,}

\affiliation[a]{Instituto de F\'{i}sica Te\'orica UAM-CSIC, Nicol\'as
  Cabrera 13-15,  Cantoblanco, E-28049 Madrid, Spain}
\affiliation[b] {Departamento de Física Teórica, Universidad Autónoma de Madrid, Francisco Tomás y Valiente 7, Módulo 15, Cantoblanco, E-28049 Madrid, Spain }

\emailAdd{jorge.dasilva@uam.es}
\emailAdd{margarita.garcia@uam.es}

\abstract{ We study, by means of numerical methods, new $SU(N)$ self-dual instanton solutions on $\mathbf{R}\times \mathbf{T}^3$ with fractional topological charge $Q=1/N$. They are obtained on a box with twisted boundary conditions with a very particular choice of twist: both the number of colours and  the 't Hooft $\mathbf{Z}_N$ fluxes piercing the box are taken within the Fibonacci sequence, i.e. $N=F_n$ (the $nth$ number in the series) and $|\vec m| = |\vec{k}|=F_{n-2}$. Various arguments based on previous works and in particular on ref.~\cite{Chamizo:2016msz}, indicate that this choice of twist avoids the breakdown of volume independence in the large $N$ limit. These solutions become relevant on a Hamiltonian formulation of the gauge theory, where they represent vacuum-to-vacuum tunneling events lifting the degeneracy between electric flux sectors present in perturbation theory.   We discuss the large $N$ scaling properties of the solutions and evaluate various gauge invariant quantities like the action density or Wilson and Polyakov loop operators.
}

\preprint{%
{
IFT-UAM/CSIC-22-89
}}

\begin{document}

    \maketitle
    
    \section{Introduction}
    \label{s:introduction}
    The study of $SU(N)$ gauge fields on the hypertorus has a long history.  It was initiated by 't Hooft~\cite{tHooft:1979rtg,tHooft:1980kjq,tHooft:1981sps} who,     using the invariance of gauge potentials under the
center of the gauge group, worked out the topological structure of $SU(N)/\mathbf{Z}_N$ gauge fields on $\mathbf{T}^4$. He pointed out the appearance of new topological classes parameterized by an antisymmetric tensor of integers, the so-called twist tensor ($n_{\mu \nu}$), defined modulo $N$.
From a physical point of view, the spatial part of the twist, with components $ m_i =\epsilon_{ijk} n_{jk} /2$, can be interpreted in the Hamiltonian limit ($\mathbf{R} \times \mathbf{T}^3$) as the introduction of abelian $Z_N$ magnetic fluxes piercing the 3-dimensional spatial torus, while the remaining part ($k_i = n_{0i}$) is dual to the electric flux, a vector of quantum numbers characterizing the physical Hilbert space on the theory.  Likewise,
't Hooft also realized that a non-trivial twist changes the way topological charge is quantized on the torus; in fact, with       
so-called non-orthogonal twists ($\vec k \cdot \vec m \ne 0$) the topological charge is fractional,
quantized in units of $1/N$~\cite{tHooft:1981nnx,vanBaal:1982ag}. Although existence of self-dual configurations with fractional topological charge was proven by Sedlacek~\cite{Sedlacek:1982cd}, the only analytic solutions obtained so far are constant curvature abelian ones which become self-dual only for particular values of the torus aspect ratios~\cite{tHooft:1981nnx,Gonzalez-Arroyo:2019wpu}.

In this paper, we introduce a new class of $SU(N)$ fractional charge self-dual solutions that are localized in time and can be interpreted in a Hamiltonian set-up as tunneling events between inequivalent flat connections on $\mathbf{T}^3$.
They are obtained numerically through a minimization procedure following the gradient flow~\cite{Narayanan:2006rf,Lohmayer:2011si,Luscher:2009eq,Luscher:2010iy} and scale nicely towards the large $N$ limit.  The first example of fractional charge solutions of this type was 
obtained for gauge group $SU(2)$ and magnetic twist $\vec m =(1,1,1)$ in refs.~\cite{GarciaPerez:1989gt,GarciaPerez:1992fj}, a case generalized later on to arbitrary number of colours~\cite{GarciaPerez:1997fq,Montero:2000mv} . Here we will analyze a different type of twist that becomes relevant in the large $N$ limit.
Our choice is a rather peculiar one in the sense that both the twist and the number of colours are taken within the Fibonacci sequence, i.e.  $N=F_n$ is itself a Fibonacci number (the $nth$ one in the series) and we take $\vec m = (0,0,F_{n-2})$~\cite{Chamizo:2016msz}. In addition, we consider an asymmetric torus geometry where two of the periods scale as $1/N$. These particular choices are dictated by considerations based on the idea of volume independence and its particular realization with twisted boundary conditions~\cite{Gonzalez-Arroyo:1982hyq,Gonzalez-Arroyo:1982hwr,Gonzalez-Arroyo:1983zev}. It has been shown that naive twists, with 't Hooft flux kept constant as $N$ is sent to infinity, lead to a breakdown of volume independence~\cite{Ishikawa,Bietenholz:2006cz,Teper:2006sp,Azeyanagi:2007su}, 
a fact also observed in the context of non-commutative gauge theories on the torus~\cite{Guralnik:2002ru}, related via Morita duality to twisted boxes~\cite{Gonzalez-Arroyo:1983zev,Schwarz:1998qj}. Several studies, both in 2+1 dimensions~\cite{Chamizo:2016msz,GarciaPerez:2013idu,GarciaPerez:2018fkj} and in 4 dimensions~\cite{Perez:2017jyq,Bribian:2019ybc}, indicate that this breakdown can be optimally avoided if the large $N$ limit is approached along the Fibonacci sequence as indicated above. We will elaborate on this point in sec.~\ref{s:general}, where we will summarize several well known facts about the formulation of gauge theories on a twisted box; for reviews on this topic the reader is referred to refs.~\cite{Gonzalez-Arroyo:1997ugn,vanBaal:2000zc,GarciaPerez:2014cmv,GarciaPerez:2020gnf}.

The paper is structured as follows. In section~\ref{s:general} we discuss gauge fields on a torus and make a historical recollection of the main lines of research where twisted boundary conditions have played a role. One of them uses the torus volume as a tunable parameter to bridge the gap between the weak and strong coupling domains of the gauge theory.
The other dwells upon the idea of volume independence in the large $N$ limit. The role of fractional instantons in this context is also highlighted. Section~\ref{s:constantcur} discusses some explicit examples of constant curvature self-dual solutions with topological charge $1/N$ that exist for our particular set up. Finally, in section~\ref{s:solutions} we present the fractional instanton solutions on $\mathbf{R} \times \mathbf{T}^3$ obtained by means of a numerical minimization approach. We end with some final remarks and conclusions. Details concerning the lattice implementation of the minimization method and other lattice observables are collected in the Appendix. 

Let us finally point out that t' Hooft fluxes lie at the core of many recent works on generalized global symmetries
and mixed anomalies, initiated with refs.~\cite{Gaiotto:2014kfa,Kapustin:2014gua,Gaiotto:2017yup}. The interest in studying gauge theories with 't Hooft fluxes has therefore been retaken in the recent literature and brings new interest to a problem with a long history behind -- we refer the interested reader to refs.~\cite{Unsal:2020yeh,Cox:2021vsa,Tanizaki:2022ngt}
for further discussion on the connection between the two approaches. 

    \section{General considerations}
    \label{s:general}
    The use of the torus size as tool to explore the dynamics of Yang-Mills theories from the perturbative regime
to the onset of confinement was advocated in the 80s starting with the work by L\"uscher~\cite{Luscher:1982uv,Luscher:1982ma}.
With twisted boundary conditions, this line of thought was explored by Gonz\'alez-Arroyo and collaborators, beginning with the perturbative analysis
of the pure Yang-Mills spectrum on a small twisted box~\cite{GonzalezArroyo:1987ycm,Daniel:1989kj} and continuing with the study of the dynamical role
played by fractional instantons in $SU(2)$ Yang-Mills theory beyond the purely semiclassical
regime~\cite{RTN:1993ilw,GarciaPerez:1993jw,Gonzalez-Arroyo:1995ynx,Gonzalez-Arroyo:1995isl,Gonzalez-Arroyo:1996eos,GarciaPerez:1997fq} -- see also refs.~\cite{Itou:2018wkm,Bribian:2021cmg} for some results on $SU(3)$.
A parallel path for periodic boundary conditions was followed by van Baal and collaborators, leading to a piece of work nicely summarized in
ref.~\cite{vanBaal:2000zc}.
 In this section, we will focus on the case of twisted boundary conditions and 
collect some well known facts and references that will allow to put in context the relevance of fractional instantons.
For further details, the interest reader can consult the original references or the review articles~\cite{Gonzalez-Arroyo:1997ugn,vanBaal:2000zc, GarciaPerez:2009mmu,GarciaPerez:2020gnf}.

We start by reviewing the Hamiltonian formulation of $SU(N)$ Yang-Mills theory on $\mathbf{R} \times \mathbf T^3$, with $\mathbf T^3$ a 3-dimensional torus 
endowed with twisted boundary conditions (TBC). 
The discussion will be focused on the particular set up that we consider for this paper: 
\begin{itemize}
\item
$SU(N)$ Yang Mills theory defined on an asymmetric torus of sizes $l_1=l_2=l/N$ and $l_3=l$. The number of colours $N$ is taken as the $nth$ integer in the Fibonacci sequence, i.e. $N=F_n$. 
\item
Twisted boundary conditions on the three-torus with chromo-magnetic flux $\vec m= (0,0,m)$. The flux $m$ is taken coprime with $N$ and equal to $F_{n-2}$. 
\end{itemize}
There are various reasons behind these particular choices that a priori may look peculiar. We briefly elaborate on each of them below and refer the reader to the series of works~\cite{Chamizo:2016msz,GarciaPerez:2013idu,GarciaPerez:2018fkj,Perez:2017jyq,Bribian:2019ybc} and the review~\cite{GarciaPerez:2014cmv} for further details.

The choice of an asymmetric torus geometry is driven by the idea of TEK reduction~\cite{Gonzalez-Arroyo:1982hyq,Gonzalez-Arroyo:1982hwr,Gonzalez-Arroyo:2010omx} and twisted volume independence  -- see ~\cite{GarciaPerez:2020gnf,GarciaPerez:2014cmv} and references therein. 
With TBC, colour and spatial degrees of freedom become entangled and, as a consequence, the torus periods 
get effectively enlarged in the twisted planes by a factor of $N$~\cite{Gonzalez-Arroyo:1982hyq,Coste:1986cb}. Therefore, with our choice of twist and torus geometry, the effective large $N$ dynamics is that of a symmetric torus of size $l^3$~\cite{GarciaPerez:2013idu}.  The parameter that controls the onset of non-perturbative effects in this set-up is $\Lambda l$, with $\Lambda$ the non-perturbative dynamical scale of the theory. This is so even in the large $N$ limit (taken at fixed value of $l$) where two of the torus periods shrink to zero. This {\it singular} large $N$ limit~\cite{Alvarez-Gaume:2001czc} is precisely the one of interest for this paper.  

The concrete choice of magnetic flux $m$ is driven as well by various considerations.  First, a non-zero flux $m$ taken coprime with $N$  guarantees that the perturbative potential has $N$ isolated gauge-inequivalent minima where the classical potential energy vanishes~\cite{GonzalezArroyo:1987ycm}. This is in sharp contrast with the case of periodic boundary conditions, with
infinitely many gauge inequivalent configurations with zero vacuum energy~\cite{Gonzalez-Arroyo:1981ckv}.  Perturbation theory in this second case turns out to be non-analytic in the coupling constant~\cite{Luscher:1982ma}, complicating considerably the analysis of the small volume dynamics. Second, the particular choice of $m$ and $N$ as integers in the Fibonacci sequence aims at avoiding 
 large $N$ phase transitions that invalidate volume independence. Those are seen to appear at intermediate volumes in non-perturbative lattice simulations  and lead to $\mathbf{Z}_N \times \mathbf{Z}_N$ symmetry breaking for twists of the form $\vec m = (0,0,m)$~\cite{Ishikawa,Bietenholz:2006cz,Teper:2006sp,Azeyanagi:2007su,GarciaPerez:2013idu}. Symmetry  breaking in the large $N$ limit may be avoided by judiciously scaling the flux with $N$~\cite{Gonzalez-Arroyo:2010omx}. The connection of TBC with non-commutative gauge theories~\cite{Gonzalez-Arroyo:1983zev,Schwarz:1998qj} also gives hints of a possible failure of volume independence.
By virtue of Morita duality, our set-up is equivalent to a non-commutative $U(1)$ gauge theory defined on an 
$l^3$ torus with dimensionless non commutativity parameter given by $\hat \theta= \barm /N$, with $\barm$ the modular inverse of $m$, 
i.e. $m \times\barm = 1$ mod $N$. 
The existence of tachyonic low momentum modes rendering the perturbative vacuum unstable and their relation to center symmetry  breaking and electric flux condensation in the commutative side of the equivalence was pointed out long ago~\cite{Guralnik:2002ru}. 
Various studies in 2+1 dimensions~\cite{Chamizo:2016msz,GarciaPerez:2013idu,GarciaPerez:2018fkj} and 4 dimensions~\cite{Perez:2017jyq,Bribian:2019ybc} indicate that the optimal sequence to approach the large $N$ limit without instabilities is to take $N$ and $m$ along the Fibonacci sequence with $N=F_n$ and $m=F_{n-2}$, the $nth$ and $nth-2$ Fibonacci numbers respectively~\cite{Chamizo:2016msz}. 
It is easy to see that with this choice $\barm= (-1)^{n} F_{n-2} $ and the non-commutativity parameter becomes $|\hat \theta | =  F_{n-2} /F_n$,
which in the large $N$ limit approaches $\varphi^{-2}$, with $\varphi = (1+\sqrt{5})/2$ the Golden Ratio. We will adopt this particular choice of twist for the rest of this paper.

Once our specific choice of twist and torus geometry has been justified, we move onto the Hamiltonian description of the gauge theory on the 3-torus with TBC. We recall that the torus effective size $l$ can be tuned to make the theory analytically tractable in the regime in which $\Lambda l<<1$. We summarize below what is known about the dynamics of the gauge theory in this small volume domain.

Twisted boundary conditions on $\mathbf{R} \times \mathbf T^3$ are implemented at the level of the gauge potential 
by imposing the following periodicity conditions:
\be
A_\mu(x + l_i \hat e_i) = \Omega_i(x) A_\mu(x)  \Omega_i^\dagger(x) + i \, \Omega_i(x) \partial_\mu \Omega_i^\dagger(x),
\ee
where $\Omega_i$ are $SU(N)$ transition matrices subject to the consistency conditions:
\be
\Omega_i (x+l_j \hat e_j) \, \Omega_j(x) = Z_{i j}\,  \Omega_j (x+ l_i \hat e_i)\,  \Omega_i (x),
\label{eq:consistency_x}
\ee 
with $Z_{i j} = \exp \{i 2 \pi n_{ij} /N\}$. 
An appropriate choice of gauge to analyze the Hamiltonian limit is $A_0=0$.  In this gauge the residual, space-dependent, gauge invariance 
can  be used to  bring the three spatial twist matrices to constant $SU(N)$ matrices satisfying the consistency conditions imposed by the magnetic twist~\cite{Groeneveld:1980tt,Ambjorn:1980sm}.
With our choice $\vec m= (0,0,m)$, this gives:
\begin{align}
\Gamma_1 \Gamma_2 &= e^{i \frac{2 \pi m}{N}} \Gamma_2 \Gamma_1,
\label{eq:twist1}\\
\Gamma_3 \Gamma_i &= \Gamma_i \Gamma_3 \text{, for }  i=1,2,
\label{eq:twist3}
\end{align}
reducing the periodicity conditions to:
\be
A_i (x + l_j \widehat{e}_j) =  \Gamma_j A_i (x)  \Gamma_j^\dagger.
\ee
A Fourier expansion consistent with this choice is given by~\cite{Gonzalez-Arroyo:1982hyq}:
\be
A_i (x_0, \vec x) = \frac{1}{\sqrt{l_1 l_2 l_3}} \sum'_{\vec p} \hat A_i (x_0, \vec p) e^{i \vec p \cdot \vec x} \widehat \Gamma (\vec p),
\ee
with:
\be
\widehat \Gamma(\vec p) = \frac{1}{\sqrt{2N}}  e^{i \alpha(\vec p) } \Gamma_1^{- \barm n_2}  \Gamma_2^{ \barm n_1},
\ee
 where momenta is quantized as $p_i= 2 \pi n_i /l$, $n_i \in \mathbf{Z}$, in all three directions, and where the
prime in the sum indicates the exclusion of the cases where both $n_1$ and $n_2$ are equal to zero modulo $N$. 
At the perturbative level, the fact that momentum is quantized in units of $l$ in the short directions (with $l_1=l_2=l/N$) is the first signal of volume independence, indicating
that the 3-dimensional box has the same effective size $l$ in all three spatial directions.
Note moreover that the fact that $\Gamma_1$ and $\Gamma_2$ span the algebra of the $SU(N)$ group, together with the consistency condition
eq.~\eqref{eq:twist3}, implies that necessarily $\Gamma_3$ has to belong to the center of $SU(N)$, i.e.
$\Gamma_3= z_3 \mathbbm{I}$, and therefore there are $N$ different choices of $\Gamma_3$ compatible with our boundary conditions.

Gauge fixing is not complete once the constant twist matrices have been chosen. Space-dependent gauge transformations subject to the periodicity condition:
\be
\Omega (\vec x+l_i \hat e_i) = \Gamma_i \Omega (\vec x) \Gamma_i^\dagger,
\label{eq:gauget}
\ee
are still allowed.
On top of those, one can as well consider so-called {\it singular gauge transformations}  satisfying:
\be
\Omega_{\vec s} (\vec x + l_i \widehat{e}_i) = e^{i \frac{2\pi s_i}{N}} \,  \Gamma_i \Omega_{\vec s} (\vec x )  \Gamma_i^\dagger \text{, } s_i \in \mathbf{Z}_N.
\label{eq:large-gt}
\ee
These transformations commute with the Hamiltonian and can be simultaneously diagonalized with it. States in the Hilbert space can therefore be parameterized by how their wave function transforms under $\Omega_{\vec s}$:
\be
\Psi_{\vec e} \left ([\Omega_{\vec s}] A\right) = e^{i\frac{ 2 \pi \vec e \cdot \vec s}{N}}\, 
\Psi_{\vec e} (A).
\label{eq:wavef}
\ee
The vector of integers $\vec e=(e_1,e_2,e_3)$, defined modulo $N$, characterizes the state and receives the name of electric flux~\cite{tHooft:1979rtg}.
There are $N^3$ different electric flux sectors and the Hamiltonian can be independently diagonalized in each of them.
With our choice of boundary conditions, all sectors different from $\vec{e}=\vec{0}$ are lifted in perturbation theory, except the ones of 
	the form $\vec e= (0,0,e_3)$, which remain degenerate with the vacuum~\cite{GonzalezArroyo:1987ycm}. 
This degeneracy is removed non-perturbatively by the effect of fractional instantons; the general 
expression for the instanton-induced energy splitting can be found in ref.~\cite{vanBaal:1984ra} -- 
see also~\cite{vanBaal:2000zc} -- and has been tested for SU(2) against numerical simulations in refs.~\cite{RTN:1993ilw,GarciaPerez:1993jw}. 
It is important at this stage to recall that the semiclassical weight, $\exp\{- N S /\lambda\}$, remains non-zero in the large $N$ 't Hooft limit for $Q=1/N$ fractional instantons with  action $NS=8\pi^2$. In what follows, we take into account that the quantity that {\it scales} adequately in the large $N$ limit (in the sense of becoming $N$-independent) is precisely $NS$, or $NQ$ for the case of the topological charge. 

In the Hamiltonian limit, these fractional instantons have a well defined interpretation as tunneling
events interpolating between two pure gauge configurations, which can be characterized in terms of the 
holonomies (the spatial Polyakov loops at $x_0=\pm \infty$). Actually, with our choice of twist, the only holonomies that can be different 
from zero are those that wind non trivially in the $x_3$ direction. Let us see how this comes about.
We start by noticing that with 
twisted boundary conditions, the appropriate definition of Polyakov loops winding once along a torus period is given by~\cite{vanBaal:1984ra}:
\be
P_i(x_0, \vec x) \equiv \frac{1}{N} \Tr \left (P\exp \left \{ -i \int_0^{l_i} d x_i A_i (x_0, \vec x) \right \} \Gamma_i \right),
\ee
a definition that combined with eqs.~\eqref{eq:large-gt} and~\eqref{eq:wavef}, immediately shows that $P_i$ transforms a state with electric flux $\vec e$ into one with flux  $\vec e+ \hat e_i$.
We continue with the observation that a gauge potential  $A_i(\vec x)=0$ is compatible with the choice of constant twist matrices and leads to holonomies given by:
\be
P(\gamma, w_1, w_2,w_3) = \frac{1}{N}  \Tr \left (\Gamma_1^{w_1(\gamma)}  \Gamma_2^{w_2(\gamma)}  \Gamma_3^{w_3(\gamma)} \right),
\ee
where $\gamma$ represents a closed curve and $w_i(\gamma)$ denotes its winding number in the $ith$ direction, defined modulo $N$.
In our particular case, this implies $P(\gamma, 0,0,1) = z_3$ and $P(\gamma, w_1,w_2,0)=0$, unless $w_1$ and $w_2$   
are both equal to zero (mod $N$).  Flat connections are therefore parameterized by the value of the holonomy in the third direction which takes
values over the $N$ roots of unity. The corresponding gauge potentials are mapped into one another by the action of singular 
gauge transformations of the form given in eq.~\eqref{eq:large-gt}, spanning a group isomorphic to $\mathbf{Z}_N$. 

With these preliminaries, we are ready to discuss the numerical approach used to obtain self-dual fractional instantons on the lattice and to characterize the solutions resulting from our study.
The Hamiltonian limit is attained through a limiting process in which we start by looking at minimal action solutions on a 4 dimensional torus and analyze their behaviour when the torus period in the time direction is sent to infinity. Fractional topological charge is enforced by introducing a twist in time, i.e.  by working with gauge potentials periodic in the time direction up to gauge transformations, i.e.:
\be
A_i(x+l_0 \hat e_0) = \Omega_0(\vec x) A_i (x) \Omega_0^\dagger (\vec x) + i \Omega_0(\vec x) \partial_i \Omega_0^\dagger (\vec x).
\ee
These boundary conditions act as the generators of singular gauge transformations if $\Omega_0$ is taken to satisfy eq.~\eqref{eq:large-gt}~\footnote{It is trivial to see that they map $A_i=0$ at $x_0=-\infty$ to $A_i=i \Omega_0(\vec x) \partial_i \Omega_0^\dagger (\vec x)$ at $x_0=+\infty$, thus interpolating between two flat connections.}. The only additional condition one has to impose is that these boundary conditions support fractional topological charge. For our choice of magnetic twist ($N=F_n$, $m=F_{n-2}$), and using the formula for the quantization of $Q$ with twisted boundary conditions~\cite{vanBaal:1982ag,vanBaal:1984ra}, one obtains:
\be
Q= \frac{1}{16\pi^2} \int d^4 x \Tr \left( F_{\mu \nu}(x)\widetilde F_{\mu \nu} (x)\right) = \nu - \frac {\vec k \cdot \vec m }{N} = \nu - \frac {F_{n-2} \, k_3}{F_n}  \text{, } \nu \in \mathbf{Z}.
\ee
It is easy to see that the minimal fractional topological charge $1/N$ is obtained by taking: $k_3 = -\barm = (-1)^{n+1} F_{n-2}$ and $\nu=(-1)^{n+1} F_{n-4}$. Therefore, we select a space-time twist: $\vec k = (0,0,-\barm)$.

Existence of solutions with minimal action within each twisted sector has been proven by Sedlacek~\cite{Sedlacek:1982cd}. However,
the only $Q=1/N$ self-dual solutions that have been obtained analytically so far are constant curvature solutions on the 4-torus~\cite{tHooft:1981nnx,Gonzalez-Arroyo:2019wpu} that become self-dual only for certain values of the torus periods. Many solutions of this type can be found using the particular properties of the Fibonacci numbers; we describe some concrete examples in the next section, but none of them covers the geometry we are are interested in. Localized, instanton-like configurations in the Hamiltonian limit have only been obtained numerically first for SU(2)~\cite{GarciaPerez:1989gt,GarciaPerez:1992fj} and then for SU(N)~\cite{GarciaPerez:1997fq,Montero:2000mv} but for different twist and torus geometry from the ones used in this paper. In section~\ref{s:solutions} we will present new localized-in-time numerical solutions compatible with our set-up, and discuss their scaling properties in the singular large $N$ limit mentioned before. As we will see, in some aspects they resemble vortex-like configurations as the ones obtained for $\mathbf{T}^2 \times \mathbf{R}^2$ in refs.~\cite{Gonzalez-Arroyo:1998hjb,Montero:1999gq,Montero:2000pb} but exhibit different large $N$ scaling.

    \section{Fibonacci construction of constant curvature solutions on $ \mathbf T^4$}
    \label{s:constantcur}
    In this section we discuss some solutions of the Yang-Mills equations of motion on $\mathbf T^4$ that have topological charge $Q=1/N$ and are compatible with our choice of twisted boundary conditions. They have constant curvature and become self-dual for certain values of the torus aspect ratios. Although the Hamiltonian limit of interest is not included in this set, we provide concrete examples that up to our knowledge were not previously known in the literature. They are obtained by applying the general construction of constant curvature solutions in the 4-torus recently presented in ref.~\cite{Gonzalez-Arroyo:2019wpu}, which generalizes the original one by 't Hooft~\cite{tHooft:1981nnx}. We start with a brief review following the notation introduced in ref.~\cite{Gonzalez-Arroyo:2019wpu} and then discuss the particular implementation when both $N$ and the 't Hooft fluxes belong to the Fibonacci sequence.

The construction starts by decomposing the number of colours as $N=N_1+N_2$ and correspondingly the twist tensor as:
\be
n_{\mu\nu}=n_{\mu\nu}^{(1)}+n_{\mu\nu}^{(2)},
\label{eq:twistd}
\ee
where the $n_{\mu\nu}^{(a)}$ are  taken as orthogonal twists in $SU(N_a)$.
Defining:
\be
\Delta_{\mu\nu} = n_{\mu\nu}^{(2)}N_1-n_{\mu\nu}^{(1)}N_2,
\label{eq:delta}
\ee
the gauge field and the field 
strength tensor  of the constant curvature solutions are given by:
\begin{align}
A_\mu(x)&=\pi\frac{\Delta_{\nu\mu}}{N l_\mu l_\nu} x_\nu T,\\
F_{\mu\nu}(x)&=2\pi\frac{\Delta_{\mu\nu}}{N l_\mu l_\nu}T,
\end{align}
where $T$ stands for the hermitian and traceless matrix:
\be
T=
	\begin{pmatrix}
	\mathbbm{I}_1/N_1	& 				0  			\\
	 0						& -\mathbbm{I}_2/N_2  	\\
	\end{pmatrix},
\ee
with  $\mathbbm{I}_a$ representing the identity matrix in $SU(N_a)$.

These solutions have constant action density and a 
topological charge:
\be
Q=\frac{\epsilon_{\mu\nu\rho\sigma}\Delta_{\mu\nu}\Delta_{\rho\sigma}}{8NN_1N_2}.
\ee
Moreover, it is easy to see that they satisfy twisted boundary conditions on $\mathbf{T}^4$ corresponding with transition matrices given by:
\be
\Omega_\mu(x)=e^{i\pi \omega_\mu(x)T}
	\begin{pmatrix}
	\Gamma_\mu^{(1)}	& 			0  		\\
	 0					& \Gamma_\mu^{(2)}  \\
	\end{pmatrix},
\label{transition}
\ee
where the constant 
$SU(N_a)$ matrices $\Gamma_\mu^{(a)}$ are twist eaters compatible with the orthogonal twists $ n_{\mu\nu}^{(a)}$, i.e. satisfying:
\be
\Gamma_\mu^{(a)}\Gamma_\nu^{(a)}=e^{2\pi i n^{(a)}_{\mu\nu}/N_a}\Gamma_\nu^{(a)}\Gamma_\mu^{(a)},
\label{eq:t.eater}
\ee
and where  $\omega_\mu(x)$  stands for the function:
\be
\omega_\mu(x)\equiv\frac{\Delta_{\mu\nu}x_\nu}{N l_\nu}.
\ee

We will now consider the case in which the only non-zero components of the twist matrix are 
$\Delta_{03}\equiv \Delta_A$ and $\Delta_{12}\equiv \Delta_B$ (any other choice can be brought to this 
form by a change of basis~\cite{Gonzalez-Arroyo:1997ugn}). The self-duality condition ($F_{\mu\nu}=\tilde{F}_{\mu\nu}$) trivially amounts in this case to
\be
\frac{l_0l_3}{l_1l_2}=\frac{\Delta_{A}}{\Delta_{B}},
\label{eq:self}
\ee
implying that self-dual constant curvature solutions of this kind exist only for fixed value of the ratio 
of torus areas in planes 03 and 12.

Given the form of the topological charge, the general minimal action solutions 
with $Q=1/N$ can be obtained by choosing $\Delta_A \Delta_B = N_1 N_2$ and \cite{Gonzalez-Arroyo:2019wpu}:
\be
\Delta_A=M_{A1}M_{A2}\text{ ; }\Delta_B=M_{B1}M_{B2}\text{ ; }
N_1=M_{A1}M_{B1}\text{ ; }N_2=M_{A2}M_{B2},
\label{eq:decomposition}
\ee
with $M_{Ai}$, $M_{Bi}$ positive integers.
Taking, without loss of generality, $N_1$ and $N_2$ to be coprime, compatibility of this choice with eq.~\eqref{eq:delta} is easily obtained by selecting~\cite{Gonzalez-Arroyo:2019wpu}:
\begin{eqnarray}
k^{(a)}&=M_{Aa}\hat k^{(a)}, \\
m^{(a)}&=M_{Ba}\hat m^{(a)},
\end{eqnarray}
satisfying the following relations:
\begin{eqnarray}
\hat k^{(2)}M_{B1}-\hat k^{(1)}M_{B2}&=&1,
\label{eq:cond_k}\\
\hat m^{(2)}M_{A1}-\hat m^{(1)}M_{A2}&=&1,
\label{eq:cond_m}
\end{eqnarray}
with $\hat k^{(a)}$ ($\hat m^{(a)}$) coprime with $M_{Ba}$ ($M_{Aa}$). It is easy to see that this choice leads to orthogonal twists of the required type.

The properties of Fibonacci numbers make them particularly suited for finding non-trivial working examples of this construction. We recall our choice of gauge parameters with the number of colours set to $N=F_n$ and with 't Hooft fluxes $\vec m = (0,0,m)$ and $\vec k =(0,0,-\barm)$, with $m=F_{n-2}$ and $\barm = (-1)^{n} F_{n-2}$. The starting point is the Honsberger identity:
\be
F_{n}=F_m F_{n-m+1}+F_{m-1}F_{n-m}.
\label{eq:Honsberger}
\ee
Setting $N=F_n$, this relation can be mapped in a straightforward way
to the decomposition $N=M_{A1} M_{B1}+M_{A2}M_{B2}$. Given the self-duality condition eq.~\eqref{eq:self}, there are many possible options. We are interested in those corresponding to the singular large $N$ limit with $l_1$ and $l_2$ scaling as $l/F_n$, for fixed $l$. Therefore we obtain:
\be
l_0 l_3= \frac{M_{A_1} M_{A_2} }{F_n^2 M_{B_1} M_{B_2}},
\ee
where, for simplicity, we have set the reference scale $l=1$.
In order to keep the area in the $03$ plane of order 1 when $n$ tends to infinity we choose:
\begin{align}
M_{A_1} M_{A_2} &= F_{n-m+1} F_{n-m}, \\
M_{B_1} M_{B_2} &= F_{m} F_{m-1}.
\end{align}
with $n-m>>1$.
One can easily work out in this case the twist decomposition, c.f. eq.~\eqref{eq:twistd}, compatible with 
the conditions given in eqs.~\eqref{eq:cond_k},~\eqref{eq:cond_m}.
The solution for the case of odd $n-m$ is provided below for completeness (for even $n-m$ one just has to exchange $A_1 \leftrightarrow A_2$, 
$B_1 \leftrightarrow B_2$ and $n_{\mu \nu}^{(1)} \leftrightarrow n_{\mu \nu}^{(2)}$). One takes:
\begin{eqnarray}
M_{B1}=F_m\text{ ; }M_{B2}=F_{m-1}\text{ ; }
M_{A1}=F_{n-m+1}\text{ ; }M_{A2}=F_{n-m},
\end{eqnarray}
with  chromo-electric flux given by\footnote{We use $F_{-m} = (-1)^{m+1} F_m$ to extend the definition of Fibonacci numbers to negative integers.}:
\begin{align}
\hat k^{(1)} &= (-1)^{n+1} F_{m-2}, \\
\hat k^{(2)} &= (-1)^{n+1} F_{m-3},
\end{align}
and magnetic flux by:
\begin{align}
\hat m^{(1)} &=  F_{n-m-1}, \\
\hat m^{(2)} &=  F_{n-m-2}. 
\end{align}
 
Let us study in more detail the possible values of the area in the $03$ plane arising from this particular choice.  
We can express Fibonacci numbers by using the well-known Binet's formula:
\be
F_n=\frac{1}{\sqrt{5}}\left(\varphi^n-(\psi)^n \right),
\ee
where $\varphi$ stands for the Golden Ratio:
\be
\varphi = \frac{1+\sqrt{5}}{2},
\ee
and $\psi\equiv-\varphi^{-1}$. In the limit $n\rightarrow\infty$,  Fibonacci numbers 
scale as $\sqrt{5}F_n=\varphi^n$, leading, for $n-m$ large, to:
\be
l_0 l_3= \frac{F_{n-m+1} F_{n-m}} {F_n^2 F_{m} F_{m-1}} \, \,  \xrightarrow[n-m \rightarrow \infty] \, \,   l_0l_3=\frac{\varphi^{1-2m}} {F_{m} F_{m-1}}
\label{eq:fibonacci-cc}
\ee
with the largest value of the area in the $03$ plane obtained for $m=2$.
Choosing $l_3=1$, this leads to a maximum value of $l_0=\varphi^{-3}\sim0.236$. If instead one fixes $l_0=l_3$, 
the maximal length is $l_0=l_3=\varphi^{-3/2}\sim 0.486$. 
Clearly under this construction one cannot approach the Hamiltonian limit ($l_0 \rightarrow \infty$, $l_3=1$).
In the next section we describe solutions obtained in that limit by using a numerical approach. We analyze first the case in which $l_0=l_3=1$ and then
look at what happens when the time extent is increased, keeping $l_3=1$ fixed. As we will see, in approaching the Hamiltonian limit, the constant curvature configuration is deformed developing a localized time-profile with tails that tend to flat connections at $x_0=\pm\infty$.  

We end this section by pointing out that by using relations between Fibonacci numbers, one can construct other type of constant curvature solutions. Taking for instance $l_1=l_2=l_3=1/F_n$, constant self-dual solutions exist for any value of $l_0$ satisfying:
\be
l_0 = \frac{F_{n-m+1} F_{n-m}} {F_n F_{m} F_{m-1}},
\ee
with arbitrary $m$. Taking for instance the decomposition $n=4m+q$ and sending $n$ to infinity by taking $m$ to infinity, one obtains:
\be
l_0 = \sqrt{5} \varphi^{q+2}
\ee
which can be made arbitrarily large by sending $q$ to infinity.
A numerical study of some $SU(N)$ fractional instanton solutions in this geometry has been done in ref.~\cite{Montero:2000mv}. The choice of twists and geometry taken in that work differ from ours but the large $N$ scaling properties of some of the solutions agree with the ones to be described in the next section.

    \section{Fractional  instantons  on $\mathbf R \times \mathbf T^3$ }
    \label{s:solutions}
    In this section we analyze a new type of $SU(N)$ instanton configurations with fractional topological charge $Q=1/N$ compatible with our choice of twisted boundary conditions.  They are obtained on the lattice as classical solutions of the lattice equations of motion. In the continuum limit, they lead to self-dual configurations defined on a 4-dimensional torus of periods
$l_0= s l$, $l_1=l_2=l/N$ and $l_3=l$. By changing $s$ we analyze their approach to the Hamiltonian limit: $\mathbf R \times \mathbf T^3$. As we will see, the resulting configurations are localized in the $03$ plane with a characteristic size of order $l$ and interpolate between two flat connections at $x_0 \rightarrow \pm \infty$, with holonomies dictated by the choice of 't Hooft fluxes, as discussed in section~\ref{s:general}.    

The solutions are obtained by means of a minimization procedure based on the gradient flow~\cite{Narayanan:2006rf,Lohmayer:2011si,Luscher:2009eq,Luscher:2010iy}.   
For this purpose, we discretize the SU(N) gauge theory on a $s NL  \times L^2 \times NL  $ lattice corresponding to the continuum torus mentioned above, with $l = N L a$ ($a$ the lattice spacing). 
Twisted boundary conditions have been implemented in the usual way:  on a periodic lattice with a modified action obtained by multiplying lattice plaquettes by the appropriate center-element twist factors~\cite{Groeneveld:1980tt,Groeneveld:1980zx}. These factors have been chosen so as to impose TBC with: $\vec m = (0,0,m)$ and $\vec k =(0,0,-\barm)$, with $m=F_{n-2}$ and $\barm = (-1)^{n} F_{n-2}$ for $SU(F_n)$. 
 
\begin{table}[ht]
\small
\centering
\begin{tabular}{rrrrccccc}
  \toprule
$N$ &   $m$ & $\barm$  &  $L$ & $s$ & $N S/8\pi^2$ & $N S_E/8\pi^2$  &$N S_B/8\pi^2$ & $N Q$ \\
 \midrule
3 & 1 &  1 & 6  & 1 & 1.000185 & 0.502579 & 0.497606 & 0.965560 \\
3 & 1 &  1 & 8  & 1 & 1.000062 & 0.501413 & 0.498650 & 0.980676 \\
3 & 1 &  1 & 10 & 1 & 1.000026 & 0.500892 & 0.499134 & 0.987648 \\
3 & 1 &  1 & 20 & 1 & 0.999995 & 0.500216 & 0.499779 & 0.996917 \\
 \midrule
3 & 1 &  1 & 4 & 2 & 1.000766 & 0.505760 & 0.495006 & 0.921408 \\
3 & 1 &  1 & 6 & 2 & 1.000186 & 0.502393 & 0.497793 & 0.965320 \\
3 & 1 &  1 & 8 & 2 & 1.000063 & 0.501309 & 0.498754 & 0.980542 \\
 \midrule
5 & 2 &  -2 & 4  & 1 & 1.000092 & 0.501876 & 0.498216 & 0.974790 \\
5 & 2 &  -2 & 6  & 1 & 1.000019 & 0.500814 & 0.499205 & 0.988820 \\
5 & 2 &  -2 & 12 & 1 & 1.000007 & 0.500203 & 0.499804 & 0.997209 \\
 \midrule
5 & 2 &  -2 &  4 & 2 & 1.000093 & 0.501714 & 0.498379 & 0.974579 \\
5 & 2 &  -2 &  6 & 2 & 1.000019 & 0.500743 & 0.499277 & 0.988727 \\
5 & 2 &  -2 & 12 & 2 & 0.999994 & 0.500179 & 0.499815 & 0.997185 \\
 \midrule
8 & 3 & 3 & 2 & 1 & 1.000233 & 0.503091 & 0.497142 & 0.959275 \\
8 & 3 & 3 & 4 & 1 & 1.000017 & 0.500734 & 0.499283 & 0.989868 \\
8 & 3 & 3 & 6 & 1 & 1.000002 & 0.500322 & 0.499680 & 0.995501 \\
8 & 3 & 3 & 8 & 1 & 1.000002 & 0.500181 & 0.499821 & 0.997470 \\
 \midrule
8 & 3 & 3 & 2 & 2 & 1.000235 & 0.502842 & 0.497393 & 0.958953 \\
8 & 3 & 3 & 4 & 2 & 1.000016 & 0.500673 & 0.499344 & 0.989788 \\
8 & 3 & 3 & 6 & 2 & 1.000001 & 0.500294 & 0.499706 & 0.995466 \\
 \midrule
13& 5 &  -5 & 2 & 1 & 1.000036 & 0.501109 & 0.498926 & 0.984834 \\
13& 5 &  -5 & 4 & 1 & 1.000001 & 0.500284 & 0.499717 & 0.996187 \\
\midrule
13& 5 &  -5 & 2 & 2 & 1.000036 & 0.501015 & 0.499021 & 0.984712 \\
13& 5 &  -5 & 4 & 2 & 0.999999 & 0.500246 & 0.499752 & 0.996185 \\
\midrule
21& 8 &  8 & 2 & 1 & 1.000005 & 0.500420 & 0.499586 & 0.994169 \\
\midrule
21& 8 &  8 & 2 & 2 & 1.000005 & 0.500384 & 0.499621 & 0.994123 \\
\bottomrule
\end{tabular}
\caption{Value of total action ($S$) and its electric ($S_E$) and magnetic ($S_B$) parts in units of $8 \pi^2/N$. We also give the value of $N$ times the topological charge ($NQ$). Results are given for all our lattice configurations representing a continuum torus with periods: $l_0=sNLa$, $l_1=l_2 = L a$ and $l_3=NLa$, with $a$ the lattice spacing. Quantities are reported with an error of 1 in the last significant digit, taking into account the changes in the last iteration of the flow minimization.}
\label{tab:conf}
\end{table} 

\begin{figure}[t]
\begin{subfigure}{.9\textwidth}
  \centering
  \includegraphics[width=\linewidth]{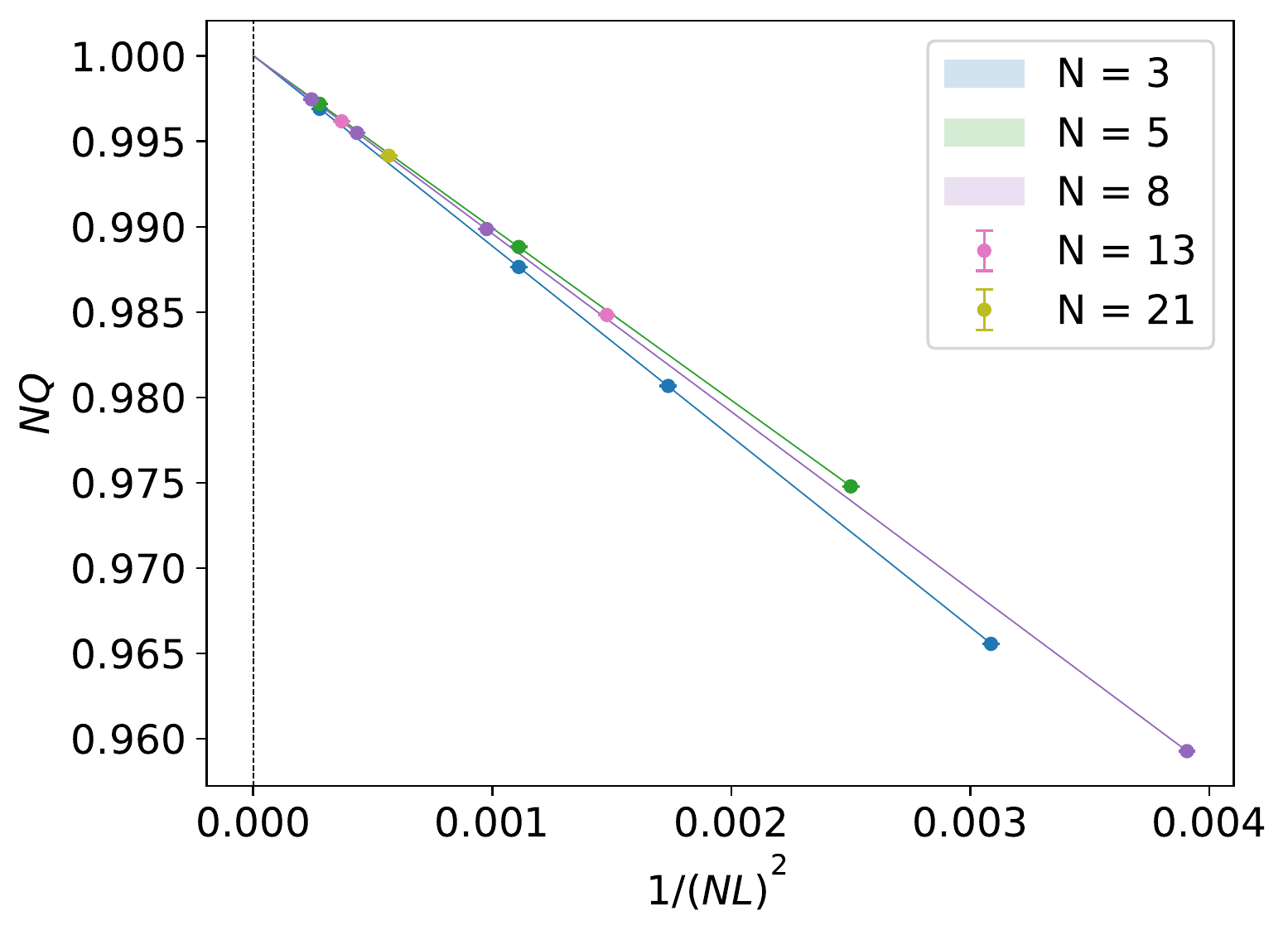}
\end{subfigure}
\caption{Continuum extrapolation of ($N$-times) the topological charge: $NQ$. Data is extracted from table~\ref{tab:conf} and is extrapolated to the continuum with $a^2= 1/(NL)^2$ corrections.}
\label{fig:extrap}
\end{figure}

The set of lattices employed in our study is given in table~\ref{tab:conf}. We have analyzed various gauge groups, with $N=3,5,8,13,21$ in the Fibonacci sequence. Setting the scale by fixing $l = 1$, the continuum limit is taken by sending the lattice spacing $a=1/LN$ to zero
 at fixed value of $N$, i.e. by sending $L$ to infinity.
 In all cases, except for $N=21$, we have results at various values of $L$ that allow to explore the approach to the continuum.
 Further details on the numerical implementation of the minimization procedure as well as the definition of the lattice observables used to measure the topological charge and action density profiles are presented in the Appendix, here we focus on describing the properties of the solutions and the scaling in the singular large $N$ limit.  

We have determined the topological charge ($Q$) and the total action ($S$) of the minimum action configuration obtained in each case.
In addition, and in order to test the self-duality of the solutions, we have separately computed the electric ($S_E$) and magnetic ($S_B$) contributions to the total action. The resulting values of $S$, $S_E$, $S_B$, in units of $8\pi^2/N$, as well as the topological charge multiplied by $N$ are given in table~\ref{tab:conf}. In all cases the results are very close to those expected in the continuum for a self-dual solution with fractional topological charge $1/N$ ($NQ=1$, $NS/(8\pi^2)=1$,  $NS_E/(8\pi^2)=NS_B/(8\pi^2)=0.5$), an indication of the smallness of lattice artefacts and the smoothness of the solutions obtained.
We can also use the results at various values of the lattice spacing to obtain a continuum extrapolation of these quantities, and we show in fig.~\ref{fig:extrap} the one of $NQ$.  A quadratic extrapolation in the lattice spacing gives values that agree with the continuum result up to one part in  $10^4$. We point out that anti-instanton configurations with $Q=-1/N$ can also be obtained under time reversal, which at the level of boundary conditions is implemented by changing the sign of the electric twist $\barm$. In the Hamiltonian limit both configurations contribute to the semi-classical partition function.

In addition to these global quantities, we have also studied several gauge invariant observables that can be used to characterize the solution. We start with a description of action density profiles and end with the analysis of Polyakov and Wilson loops. Our main interest will be in analyzing the large $N$ scaling of the solutions. 

\subsection{Action density profiles}
\begin{figure}[t]
\begin{subfigure}{.9\textwidth}
  \centering
  \includegraphics[width=\linewidth]{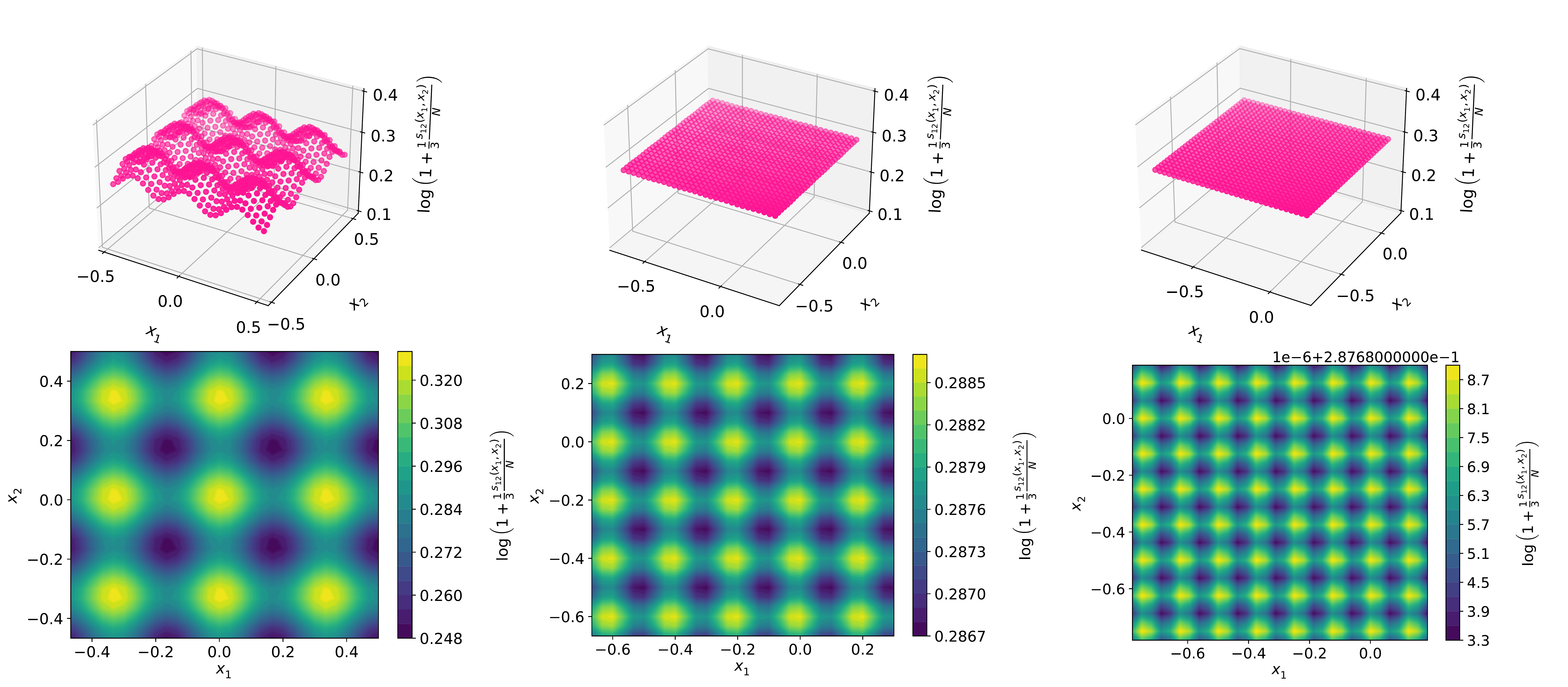}
\end{subfigure}
\caption{We display  as a function of $x_1/l$ and $x_2/l$ the profiles obtained by integrating the action density over $x_0$ and $x_3$. The unit of length is set by taking $l=1$. Gauge groups are, from left to right: $SU(3)$, $SU(5)$ and $SU(8)$.  For readability of the plots, the quantity displayed is  $\log(1+l^2 s_{12} (x_1,x_2)/(3N))$.  In the large $N$ limit the profile approaches the one of the constant curvature solution for which $\log(1+1/3) = 0.287682$.}
\label{fig:q12}
\end{figure}

We have looked at action density profiles obtained by integrating the 4-dimensional action density (in units of $8\pi^2$) in either two or three directions: 
\begin{align} 
N s_{\mu \nu} (x_\mu,x_\nu) &\equiv \left (\prod_{\rho \ne \mu, \nu}\int_0^{l_\rho} d x_\rho \right ) \, N s(x),
\label{eq:profiles-2d}\\
N s_\mu(x_\mu)  &\equiv \left (\prod_{\rho \ne\mu} \int_0^{l_\rho} d x_\rho\right ) \, N s(x).
\label{eq:profiles-1d}
\end{align}
\begin{figure}[t]
\begin{subfigure}{.9\textwidth}
  \centering
  \includegraphics[width=\linewidth]{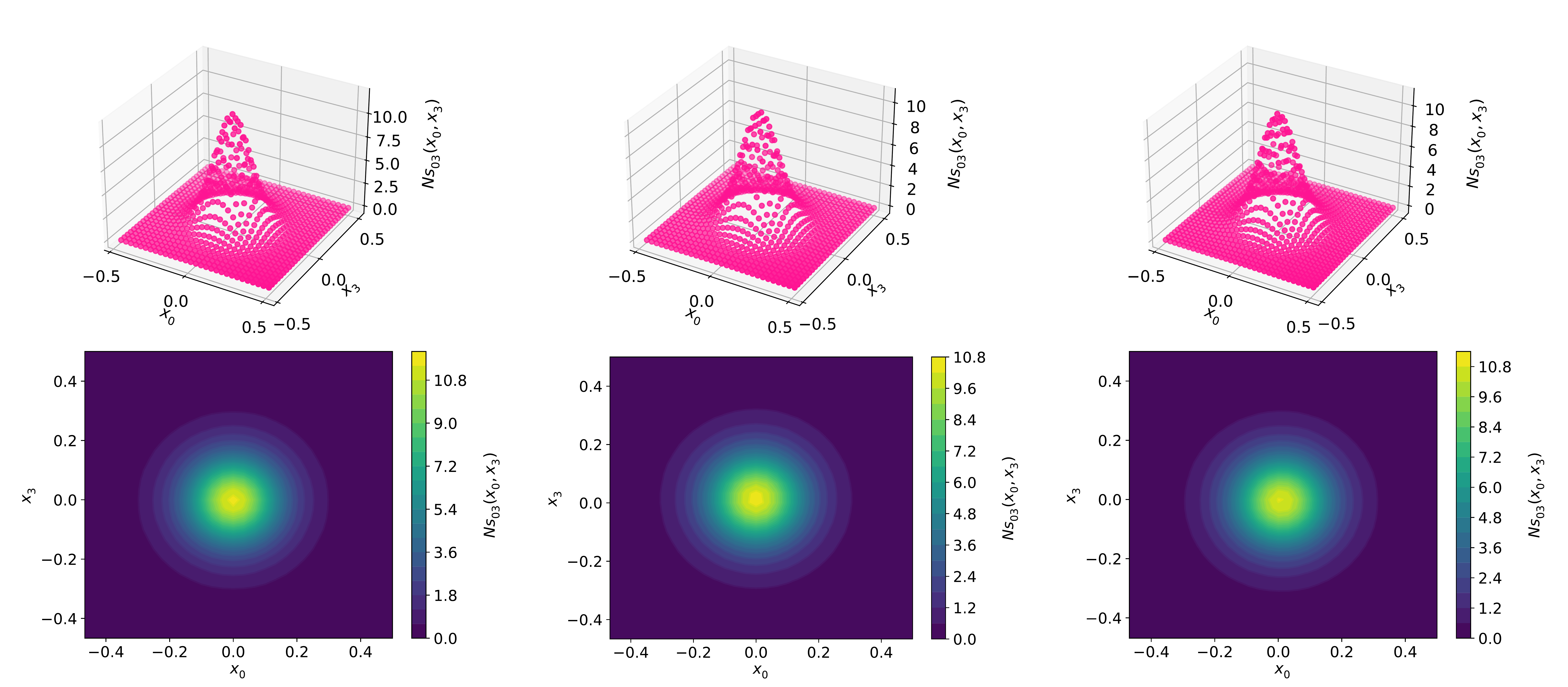}
\end{subfigure}
\caption{We display as a function of $x_0/l$ and $x_3/l$ the profiles
obtained by integrating over $x_1$ and $x_2$ the action density of the configuration with $l_0=l_3=l$ ($s=1$).
Gauge groups are, from left to right: $SU(3)$, $SU(5)$ and $SU(8)$.  }
\label{fig:s03-l0}
\end{figure}

\begin{figure}[t]
\begin{subfigure}{.9\textwidth}
  \centering
  \includegraphics[width=\linewidth]{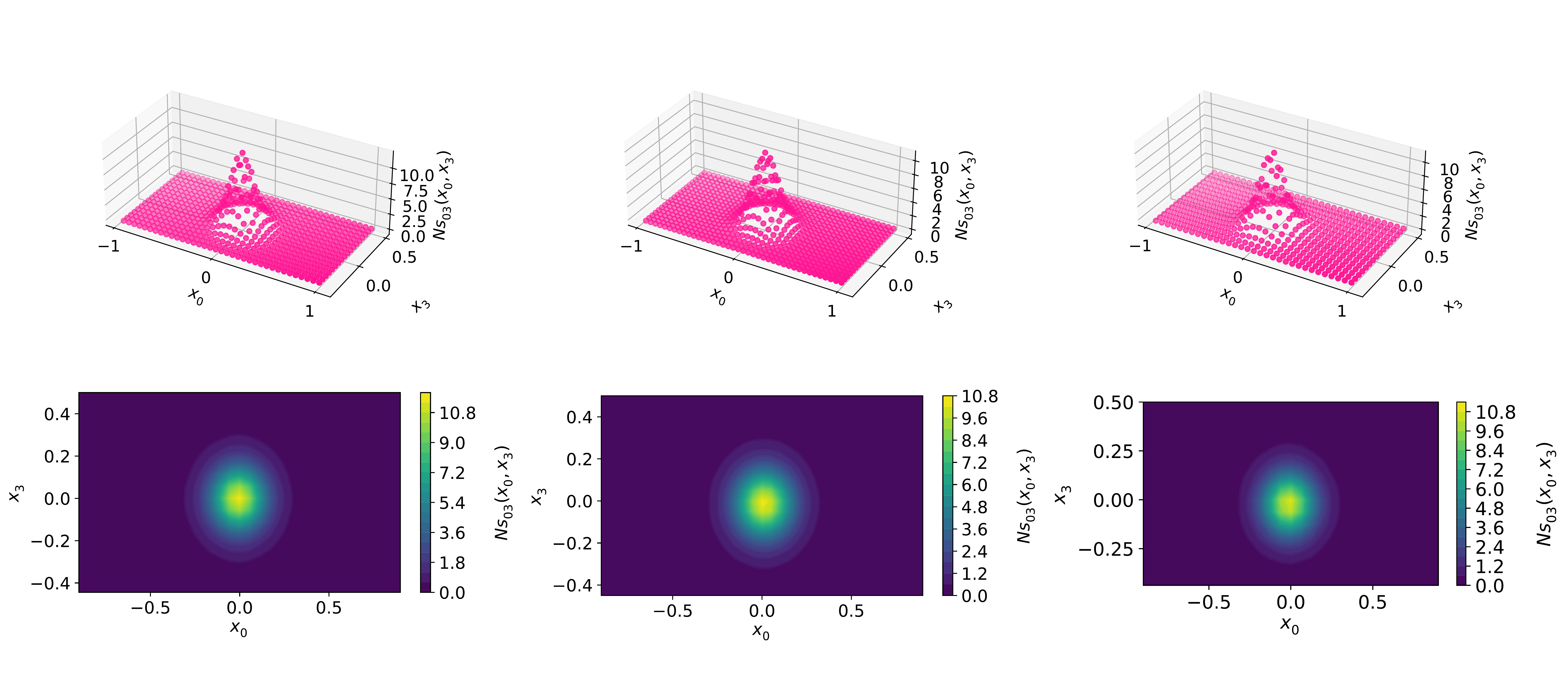}
\end{subfigure}
\caption{The same as in figure~\ref{fig:s03-l0} but for  $l_0=2l_3=2l$ ($s=2$).}
\label{fig:q03-2l0}
\end{figure}

We have introduced a factor of $N$ in the profiles in agreement with the expected large $N$ scaling discussed in section~\ref{s:general}, dictated by the fact that the $N$-independent quantity in the large $N$ limit is $N$ times the total action.
As a reference point, one can compute these quantities for the constant curvature solutions discussed in the previous section. The action density is given by $N s(x)= 1 / V$, with $V=\prod_\mu l_\mu$.  Setting $l_1=l_2=l/N$, $l_3=l$ and $l_0=\varphi^{-3}l$, as for the Fibonacci construction, we obtain the following scaling with $N$ for the dimensionless density profiles: 
\be
l^2 s_{12} (x_1,x_2)/N= 1 \text{ ; } N l^2 s_{03} (x_0,x_3)= \varphi^3 \text{ ; } N l s_0 (x_0) \sim 1.
\ee
As we will see below in more detail, our numerical solutions respect a similar scaling --
once the overall factor of $N$ is taken into account in this way, the remnant $N$ dependence of the action density profiles turns out to be very small. 
This goes in the direction of indicating that our solutions could perhaps be obtained as a {\it small} deformation of the constant curvature solution, along the lines presented in refs.~\cite{GarciaPerez:2000aiw,Gonzalez-Arroyo:2019wpu}; we will come back to this point later on.

Let us start with the analysis of two-dimensional profiles $s_{12}$, resulting from the integration of the action density along the two large directions. The dependence on  $x_1/l$ and $x_2/l$ for various gauge groups is shown in fig.~\ref{fig:q12}.  In the plot we have set the scale by taking $l=1$, a convention that will be used from now on.  In addition, we have used periodicity to extend the action density to a 
box of size $l$ in both directions. It is clear from the plots that at large $N$ these density profiles approach those of the constant curvature solutions with $l_1=l_2=1/N$. 

As for profiles depending on the two large periods, they are displayed for various gauge groups and $l_0=l_3$ or $l_0=2l_3$ in figs.~\ref{fig:s03-l0} and~\ref{fig:q03-2l0} respectively. The structure is in both cases very similar. The solutions develop a maximum in the $03$ plane with a size determined by the effective spatial torus size $l$.
Enlarging the time period by a factor of two modifies the profiles mainly in the time tails, a fact that is best appreciated by looking at the contour plots given in fig.~\ref{fig:q03-2l0}.

A more quantitative comparison between solutions for different gauge groups, can be done by looking at one-dimensional profiles, obtained by integrating the action density in all directions but one. We will focus on profiles where the action density has been integrated over space, giving therefore the time dependence of the energy.

\begin{figure}[t]
\begin{subfigure}{.48\textwidth}
  \centering
  \includegraphics[width=\linewidth]{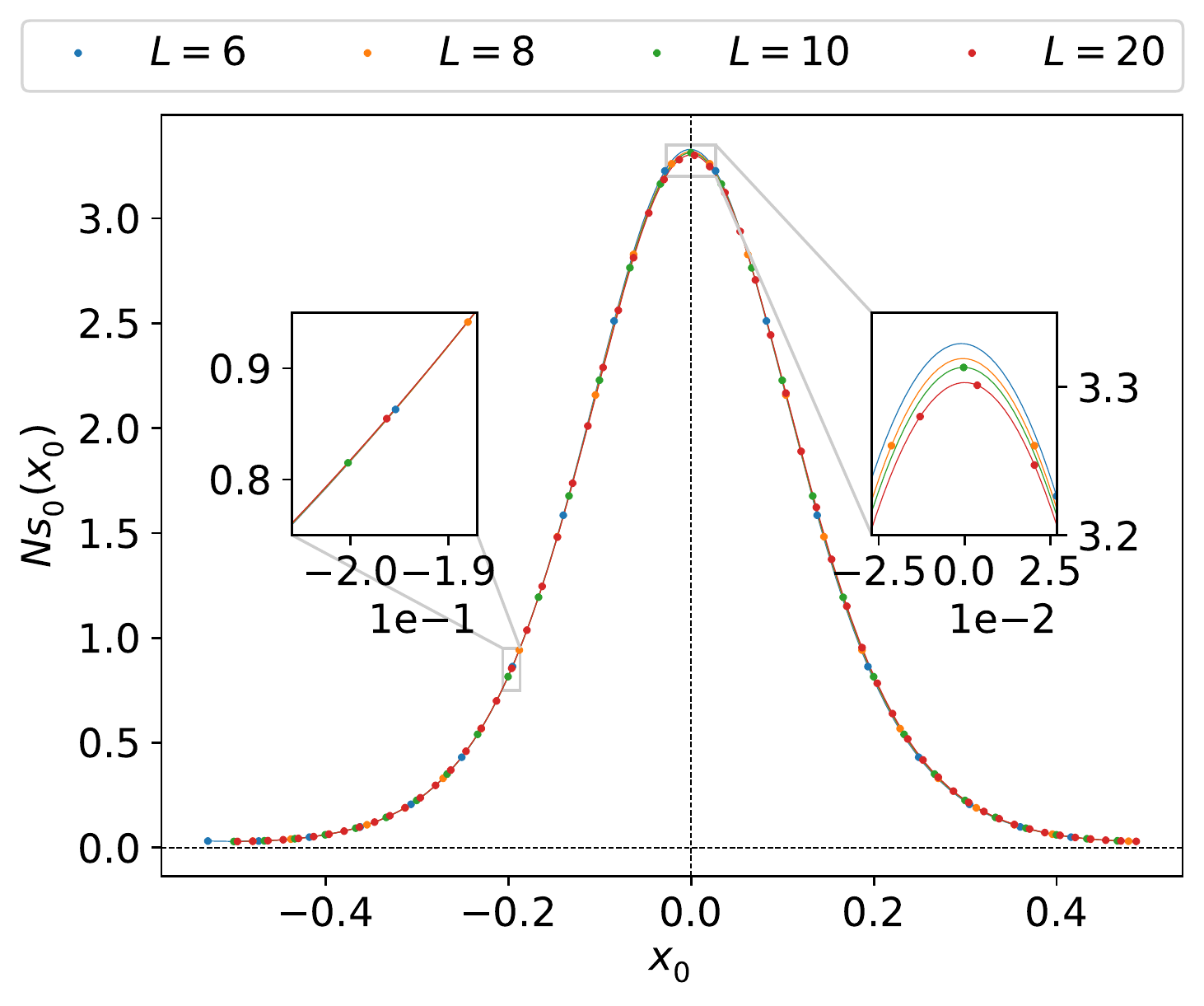}
  \caption{SU(3)}
\end{subfigure}
\begin{subfigure}{.48\textwidth}
  \centering
  \includegraphics[width=\linewidth]{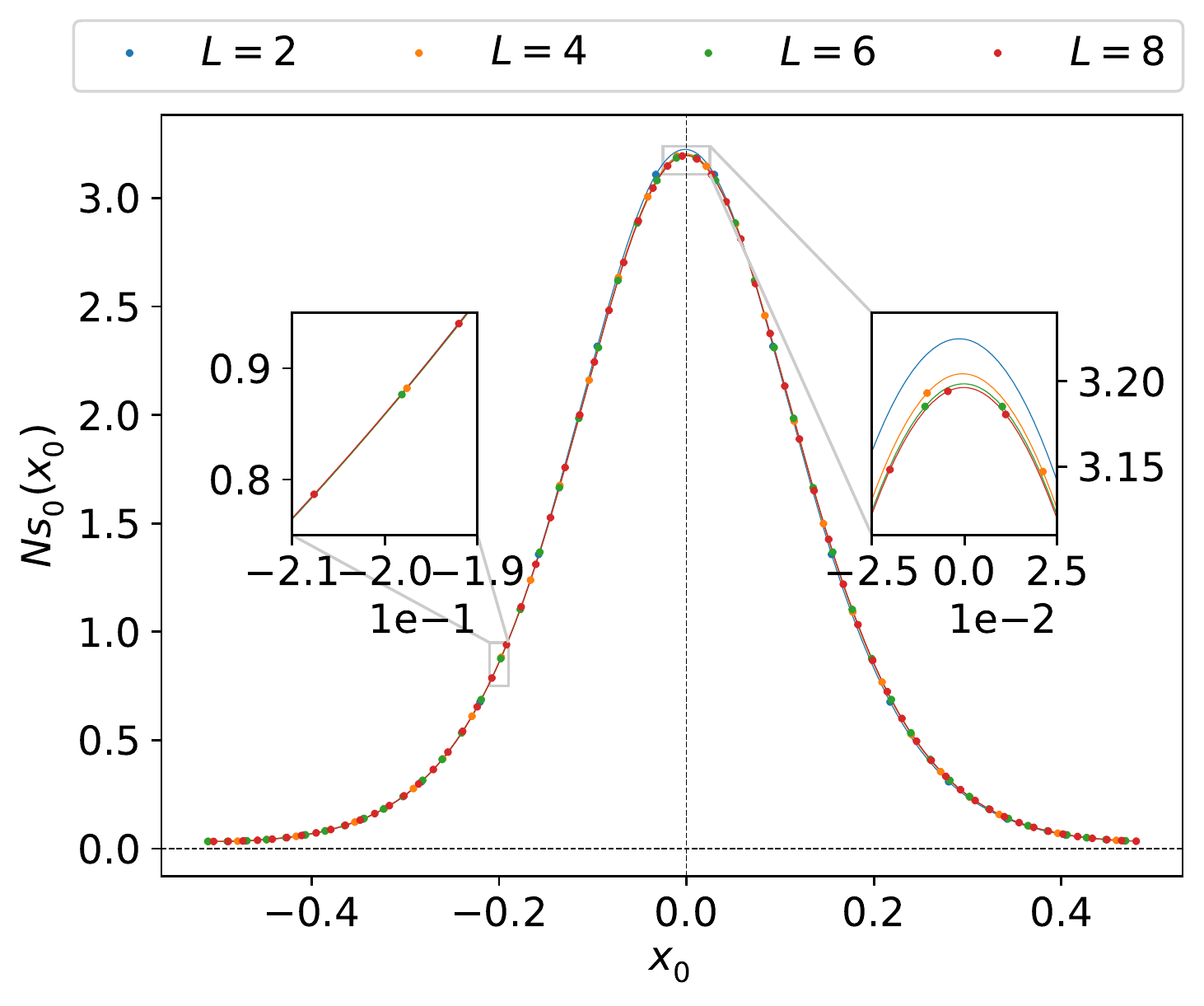}
  \caption{SU(8)}
\end{subfigure}
\caption{Continuum extrapolation of the one dimensional energy-profiles, c.f. eq.~\eqref{eq:profiles-1d} for gauge groups $SU(3)$ and $SU(8)$. The torus has been discretized on a lattice with $l_0=l_3=N L  a$ and  $l_1=l_2=L a$ with $a$ the lattice spacing. The continuum limit is taken by sending $a= l/NL$ to zero at fixed value of $l=1$, i.e. by sending $L$ to infinity.}
\label{fig:s0-cont}
\end{figure}

We first look at the continuum limit. As an example, on fig.~\ref{fig:s0-cont} we display $N l s_0(x_0)$ as a function of $x_0/l$ for gauge groups $SU(3)$ and $SU(8)$ -- similar plots are obtained for the other cases in table~\ref{tab:conf}. The plots display our results for various lattice spacings.  The continuous lines shown in each plot are interpolations to the data at fixed value of $L$, displayed to guide the eye. As already mentioned, the continuum limit is taken by sending $a=l/(LN)$ to zero. The energy-profiles (obtained using the improved clover average discussed in the Appendix) show very little dependence on the lattice spacing, indicating rather small lattice artefacts even for our smaller lattices. One important point is that discretization effects are controlled by $LN$, in such a way that even the smallest lattices with $L=2$ show a small departure from the continuum result if $N$ is large.

\begin{figure}[t]
\begin{subfigure}{.5\textwidth}
  \centering
  \includegraphics[width=\linewidth]{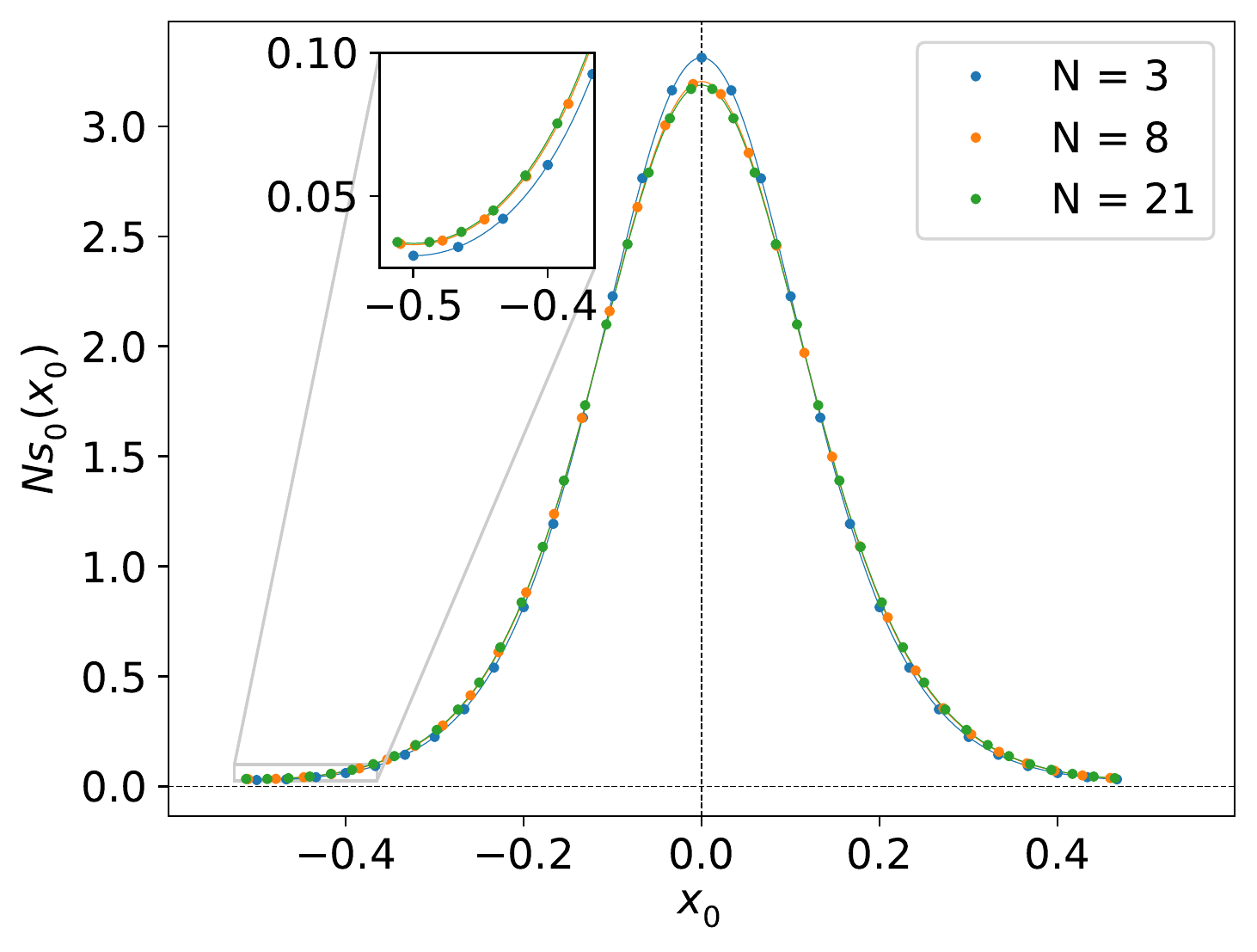}
  \caption{$s=1$}
  \label{fig:q0-variousN}
\end{subfigure}%
\begin{subfigure}{.5\textwidth}
  \centering
  \includegraphics[width=\linewidth]{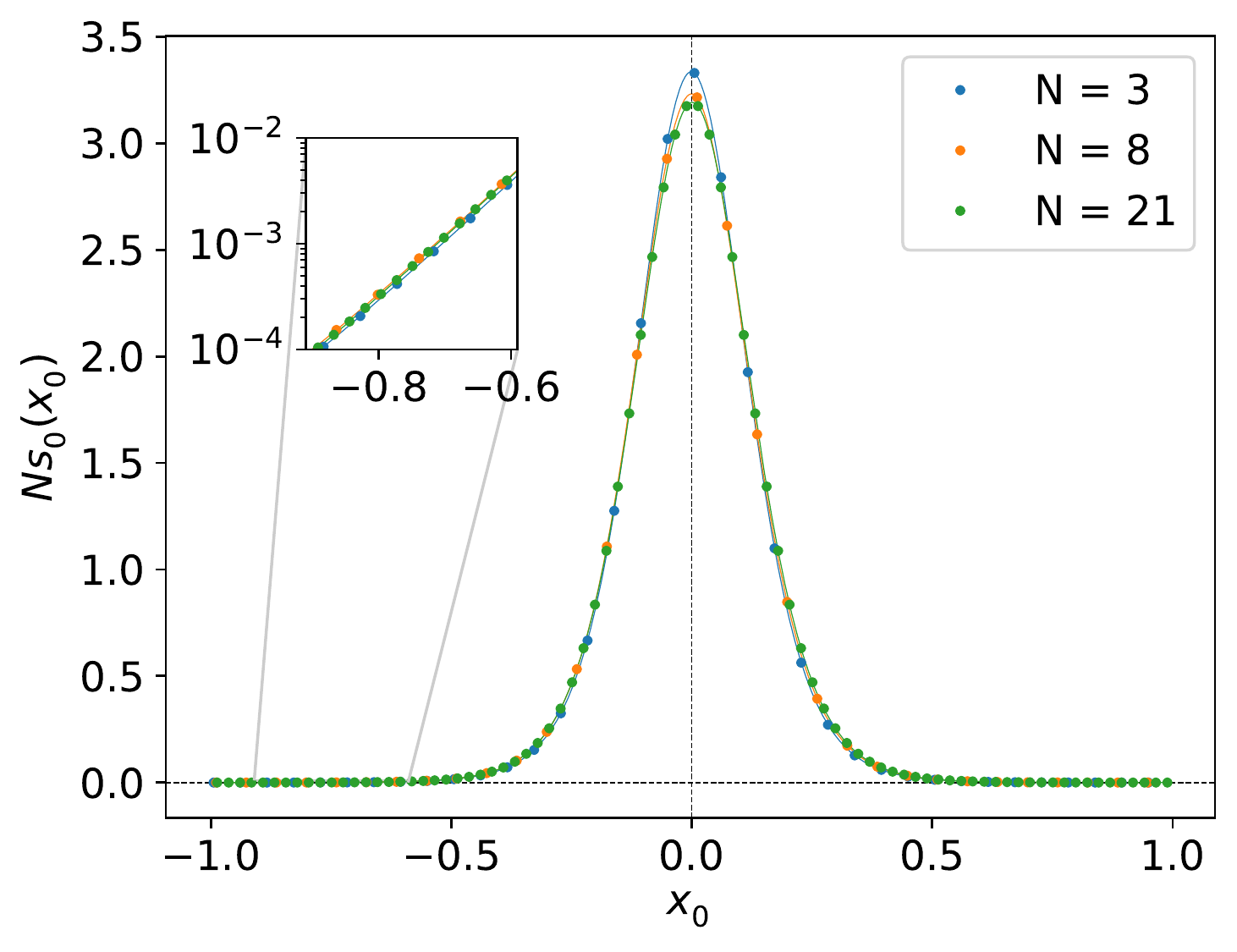}
  \caption{$s=2$}
  \label{fig:q0-variousN-2l0}
\end{subfigure}
\caption{We compare the time dependence of the energy for various gauge groups $N$ and time extents $l_0=sl$, with $s=1$ and 2. The inset in the right plot is in logarithmic ($y-$) scale to show that the decay in the tails is exponential in time.}
\end{figure}

Let us move now to the study of the $N$ dependence of the energy-profiles. 
In figure~\ref{fig:q0-variousN}, we display  $N l s_0(x_0)$ vs $x_0/l$ for the case in which $l_0=l_3=l$. Interpolating curves are displayed to guide the eye. It is clear that the $N$ dependence is rather small and the curve tends to a universal one in the large $N$ limit. 
As seen in fig.~\ref{fig:q0-variousN-2l0}, enlarging the lattice in the time direction does not change appreciably the results. The energy-profile remains localized and displays a shape very similar to the previous one except at the tails where, as shown in the inset of the plot, it decays exponentially and becomes very close to zero. This is more quantitatively measured by looking at the width at half maximum. The values obtained for all the gauge groups are given in table~\ref{tab:width}. The dependence on $N$ turns out to be very small and the same happens with the dependence on the time extent of the lattice -- compare  $w_1$ and $w_2$ corresponding to time extents differing by a factor of 2.  
Our results clearly indicate that the relevant scale parameter for the fractional instanton configurations in this geometry is given by the effective torus size $l$, set to 1 in our units. 

\begin{table}[ht]
\small
\centering
\begin{tabular}{rrrrccccc}
  \toprule
$N$ &   $m$ & $\barm$  &  $L$ &  $w_1/l$ & $w_2/l$& $w_F/l$ \\
 \midrule
3 & 1 &  1 & 6  & 0.268284 & 0.268456 & 0.222222  \\
5 & 2 & -2 & 12 & 0.283232 & 0.289695 & 0.240000  \\
8 & 3 &  3 & 6  & 0.279391 & 0.285402 & 0.234375  \\
13& 5 & -5 & 4  & 0.286707 & 0.287152 & 0.236686  \\
21& 8 &  8 & 2  & 0.281166 & 0.286412 & 0.235828  \\
\bottomrule
\end{tabular}
\caption{Value of the width at half-maximum for $l_0=l_3=l$ ($w_1$) and $l_0=2 l_3=2l$ ($w_2$) compared with the value of $l_0$ for the Fibonacci solution ($w_F$), given by eq.~\eqref{eq:fibonacci-cc} with $m=2$ and $l_3=l$.}
\label{tab:width}
\end{table} 

\begin{figure}[t]
  \centering
  \includegraphics[width=0.9\linewidth]{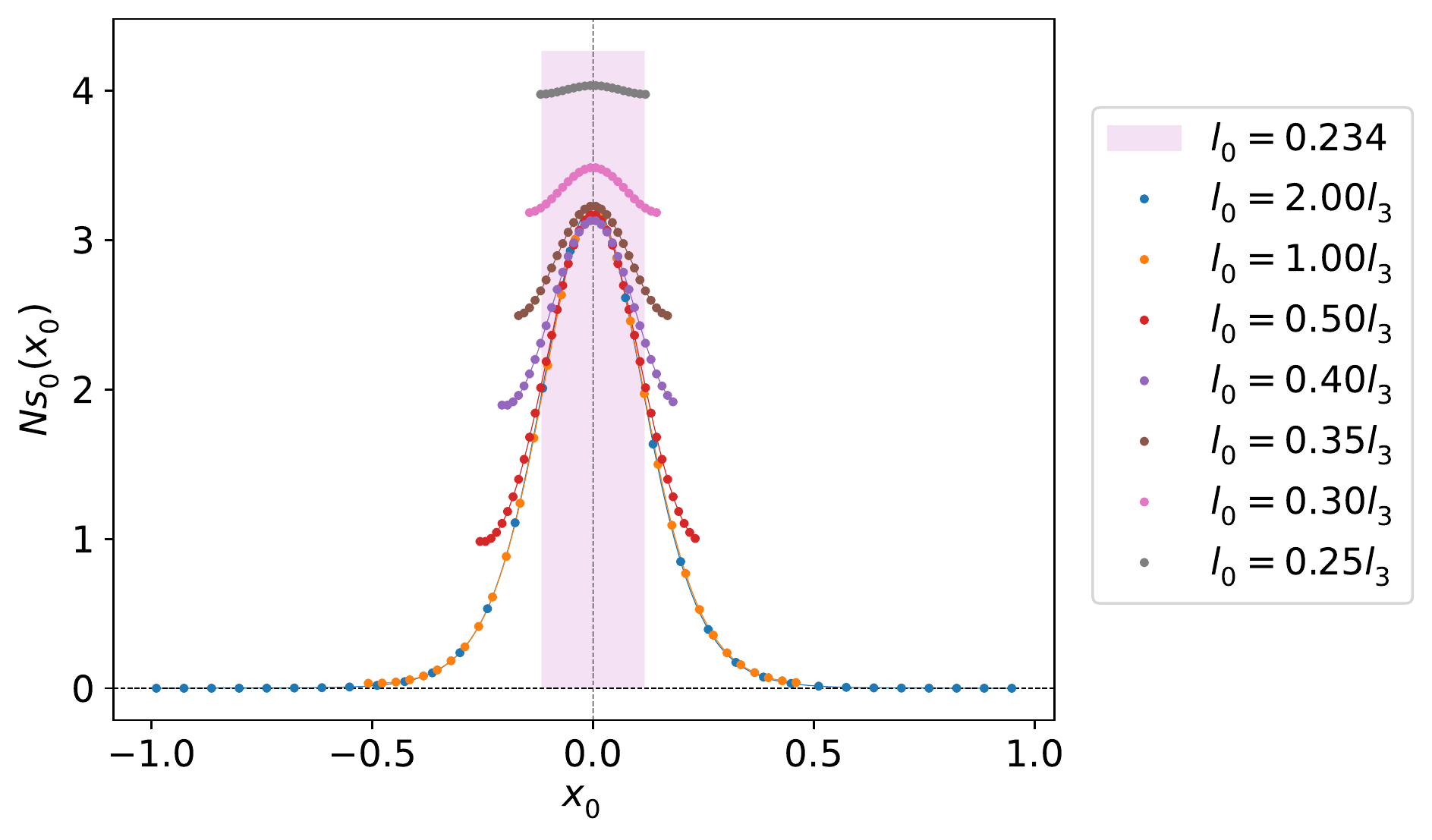}
\caption{Solutions obtained for our spatial torus geometry and various values of the time extent $l_0$ changing from the Fibonacci constant curvature case ($l_0=0.234 l $) represented by the pink box,  to the case $l_0=2l$. }
\label{fig:flat}
\end{figure}

The comparison of the energy-profile with the constant-curvature Fibonacci solution discussed in the previous section is very informative. Table~\ref{tab:width} gives the maximal value of $l_0$ for which the Fibonacci solution sharing our 3-dimensional torus geometry becomes self-dual (denoted by $w_F$). It is remarkable that this value, which tends to $w_F=0.236 l$ in the large $N$ limit, agrees surprisingly well with the instanton width. This suggests that when we distort the geometry away from the Fibonacci solution by sending $l_0$ to infinity, most of the action density remains concentrated within $w_F$. To test this hypothesis, we have generated a series of numerical solutions starting at $l_0=0.25 l$, very close to the Fibonacci case, and moving away from it in small steps. The energy-profiles for the sequence of solutions with $N=8$ are displayed in figure~\ref{fig:flat}. The change in the first steps is very fast, and beyond $l_0\sim0.5 l$ the central part of the solution remains essentially unchanged, except at the tails where the action becomes smaller for larger values of $l_0$. This is required in order to maintain the same integral within a larger region of integration.

\begin{figure}[t]
\begin{subfigure}{.49\textwidth}
  \centering
  \includegraphics[width=\linewidth]{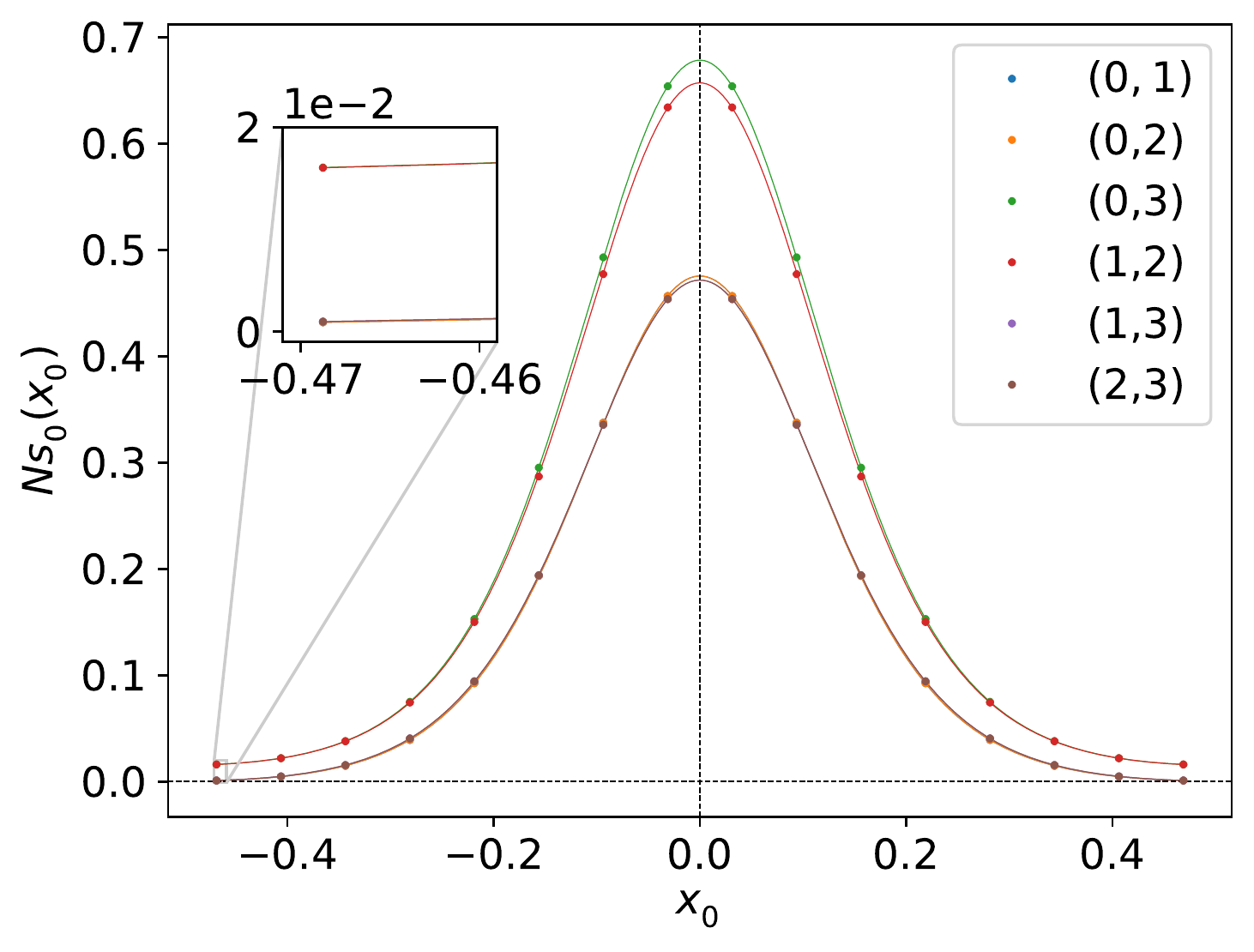}
  \caption{SU(8) with $L=2$ and $s=1$}
\end{subfigure}
\begin{subfigure}{.49\textwidth}
  \centering
  \includegraphics[width=\linewidth]{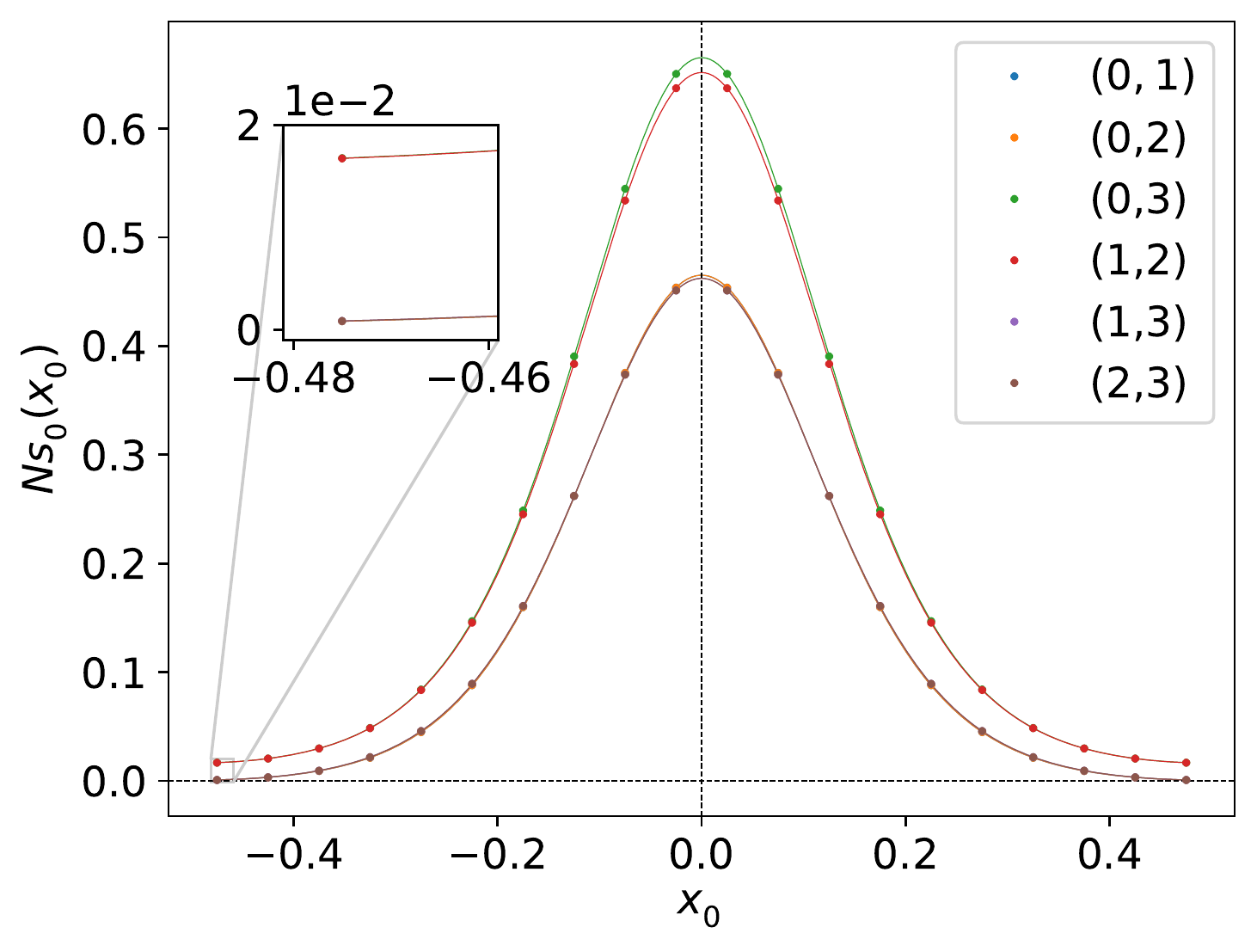}
  \caption{SU(5) with $L=4$ and $s=1$}
\end{subfigure}
\begin{subfigure}{.49\textwidth}
  \centering
  \includegraphics[width=\linewidth]{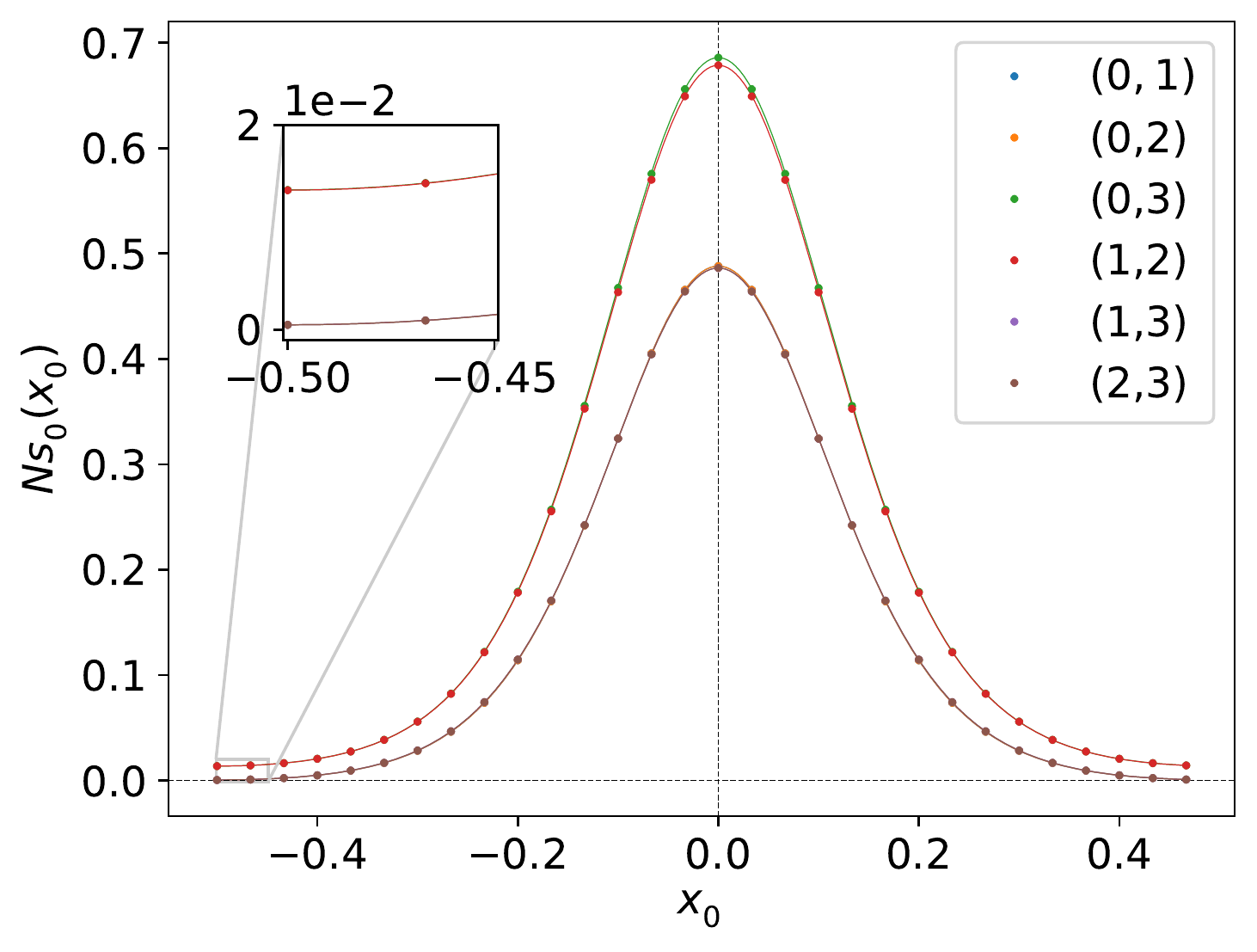}
  \caption{SU(3) with $L=30$ and $s=1$}
\end{subfigure}
\begin{subfigure}{.49\textwidth}
  \centering
  \includegraphics[width=\linewidth]{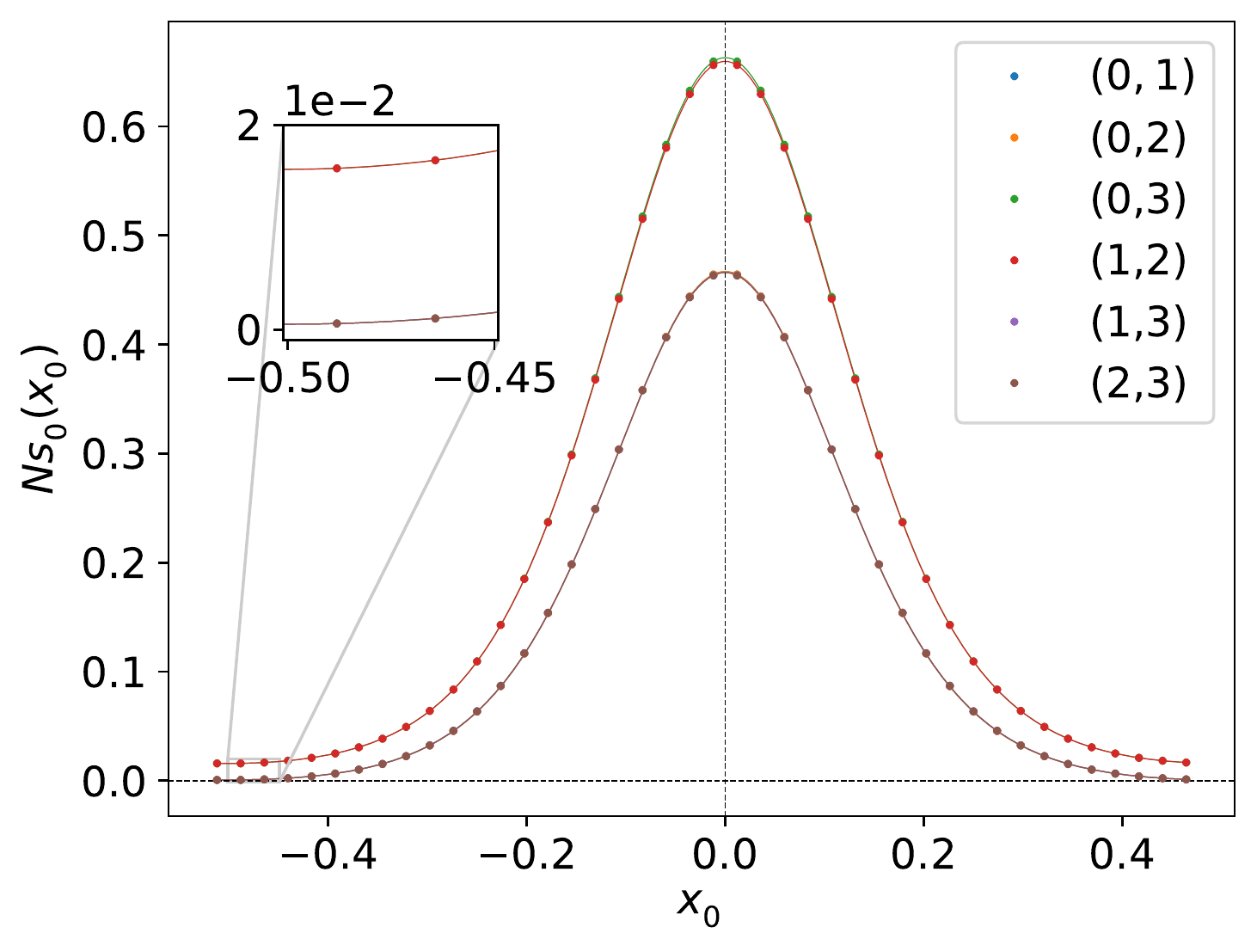}
  \caption{SU(21)with $L=2$ and $s=1$}
\end{subfigure}
\begin{subfigure}{.49\textwidth}
  \centering
  \includegraphics[width=\linewidth]{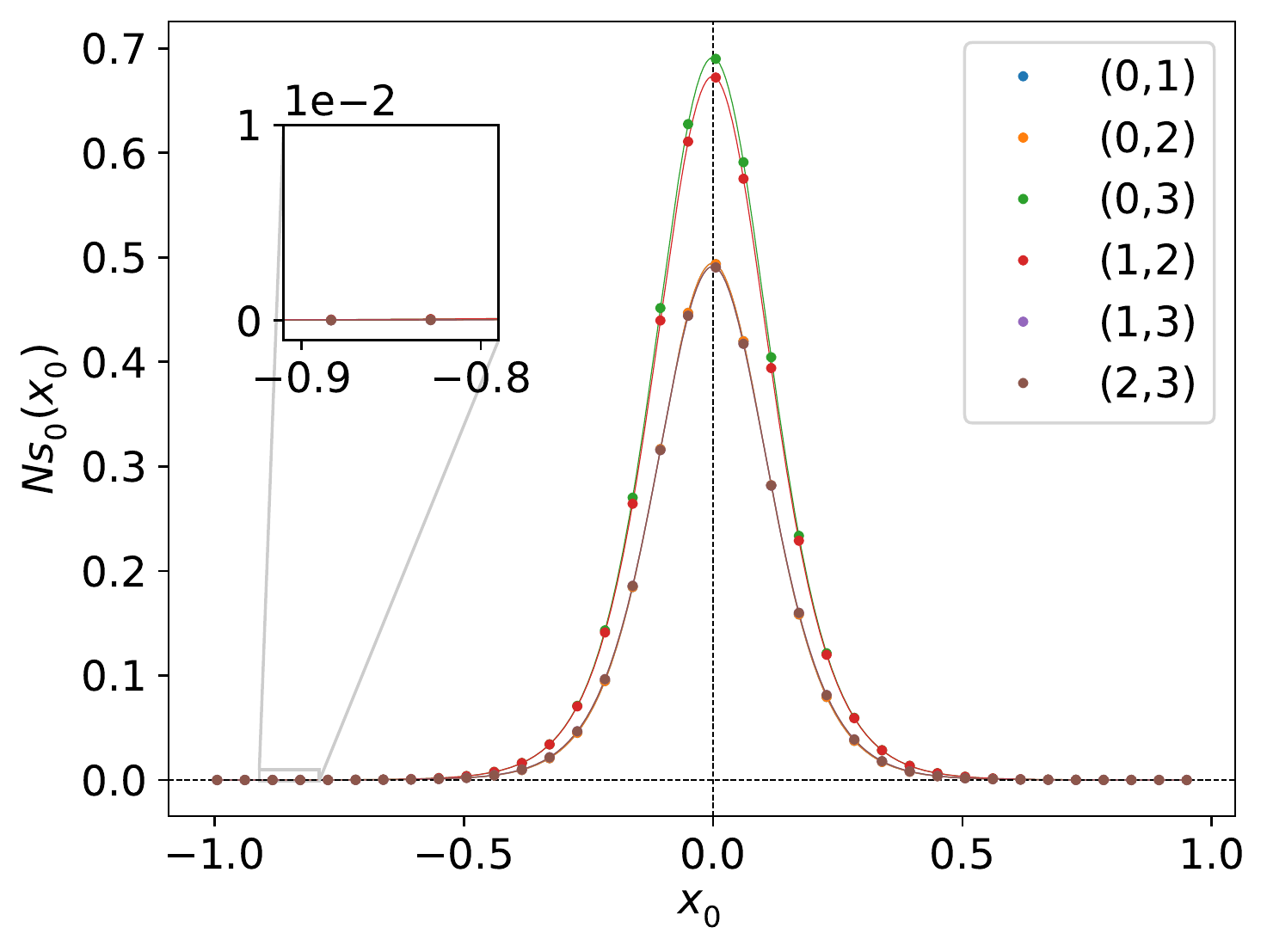}
  \caption{SU(3)with $L=6$ and $s=2$}
\end{subfigure}
\begin{subfigure}{.49\textwidth}
  \centering
  \includegraphics[width=\linewidth]{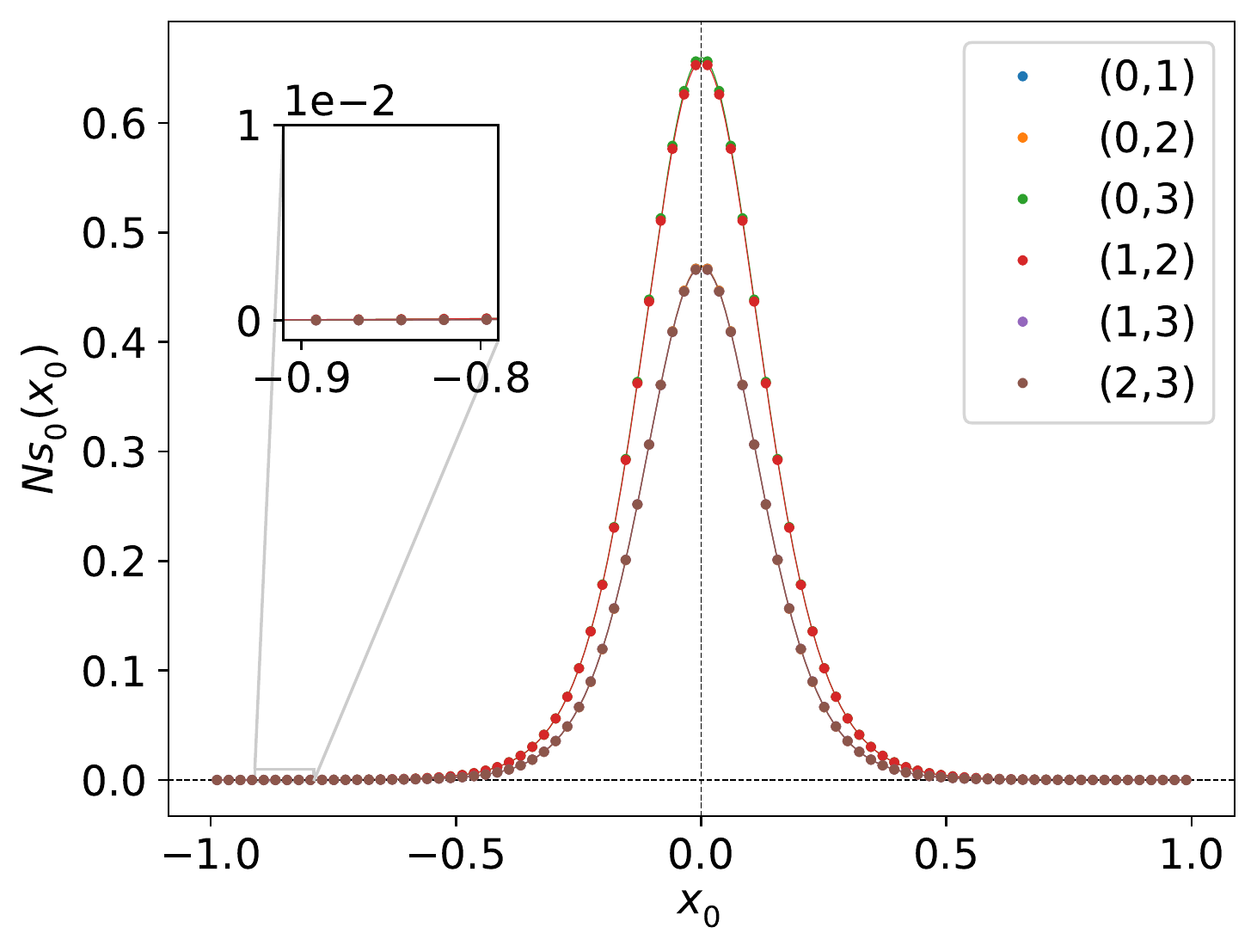}
  \caption{SU(21)with $L=2$ and $s=2$}
\end{subfigure}
\caption{Time dependence of the electric and magnetic components of the energy obtained  by integrating ${\rm Re}\Tr(F_{\mu \nu}^2)$ 
with ($\mu$, $\nu$) fixed, over the three spatial coordinates. The two largest components are 03 and 12. Components 01 and 02 are, to a very good degree, degenerate and the same happens to 23 and 13.}
\label{fig:self-dual}
\end{figure}

Finally, in order to test in a stronger way the self-duality of the solution, we have looked separately at the different components of the electric and magnetic energies. For that purpose, we have computed the spatial integral of ${\rm Re}\Tr(F_{\mu \nu}^2)$, with $\mu$ and $\nu$ fixed and analyzed their dependence on time. The results for several configurations are displayed in  fig.~\ref{fig:self-dual}. The degree of self-duality of the solution is very high and improves as we approach the continuum limit. 
Notice that for the Fibonacci solution only 
the $03$ and $12$ components of the field strength tensor are different from zero. This is no longer the case for the geometry 
we are considering, although these two components are the ones that nevertheless remain larger.

\subsection{Polyakov loops}
\label{sec:pol}
In addition to the action density, we have also analyzed other gauge invariant quantities like the Polyakov loops and the Wilson loops.

Let us start with a discussion of the Polyakov loops.
We use $P_\mu(x)$ to denote ($1/N$ times) the trace of the Polyakov loop winding the torus once in direction
$\mu$, and parameterize this quantity in terms of its modulus and phase as:
\be
P_\mu(x) = \frac{1}{N} \Tr \Big( P\exp \Big\{ -i \int_0^{l_\mu} dx_\mu A_\mu(x)\Big \}\, 
\Omega_\mu(x)\Big ) \equiv |P_\mu(x) | \, e^{i\phi_\mu(x)}.
\ee
Twisted boundary conditions reflect on the fact that Polyakov loops 
satisfy:
\be
P_\mu (x+ l_\nu \hat e_\nu) = e^{i\frac{2 \pi n_{\mu \nu}}{N}} P_\mu (x).
\ee  
Under a displacement by a torus period, the modulus is therefore periodic and the phase shifts by an amount that depends on the twist. 

As mentioned in section~\ref{s:general}, in the Hamiltonian limit ($l_0 \rightarrow \infty$) the fractional instanton interpolates between two flat connections characterized by the value of the holonomies. Those corresponding to loops in the spatially twisted plane, are maximally non trivial with: $P_i(\vec x, x_0=\pm \infty) =0$, for $i=1,2$. On the other hand, $P_3(\vec x, x_0=\pm \infty)$ is in the center of the gauge group, jumping by a factor $2 \pi n_{30}/N$ between $x_0=-\infty$ and at $x_0=+\infty$. As we will see below, our numerical configurations match very well this behaviour.

\begin{figure}[t]
\begin{subfigure}{.5\textwidth}
  \centering
  \includegraphics[width=\linewidth]{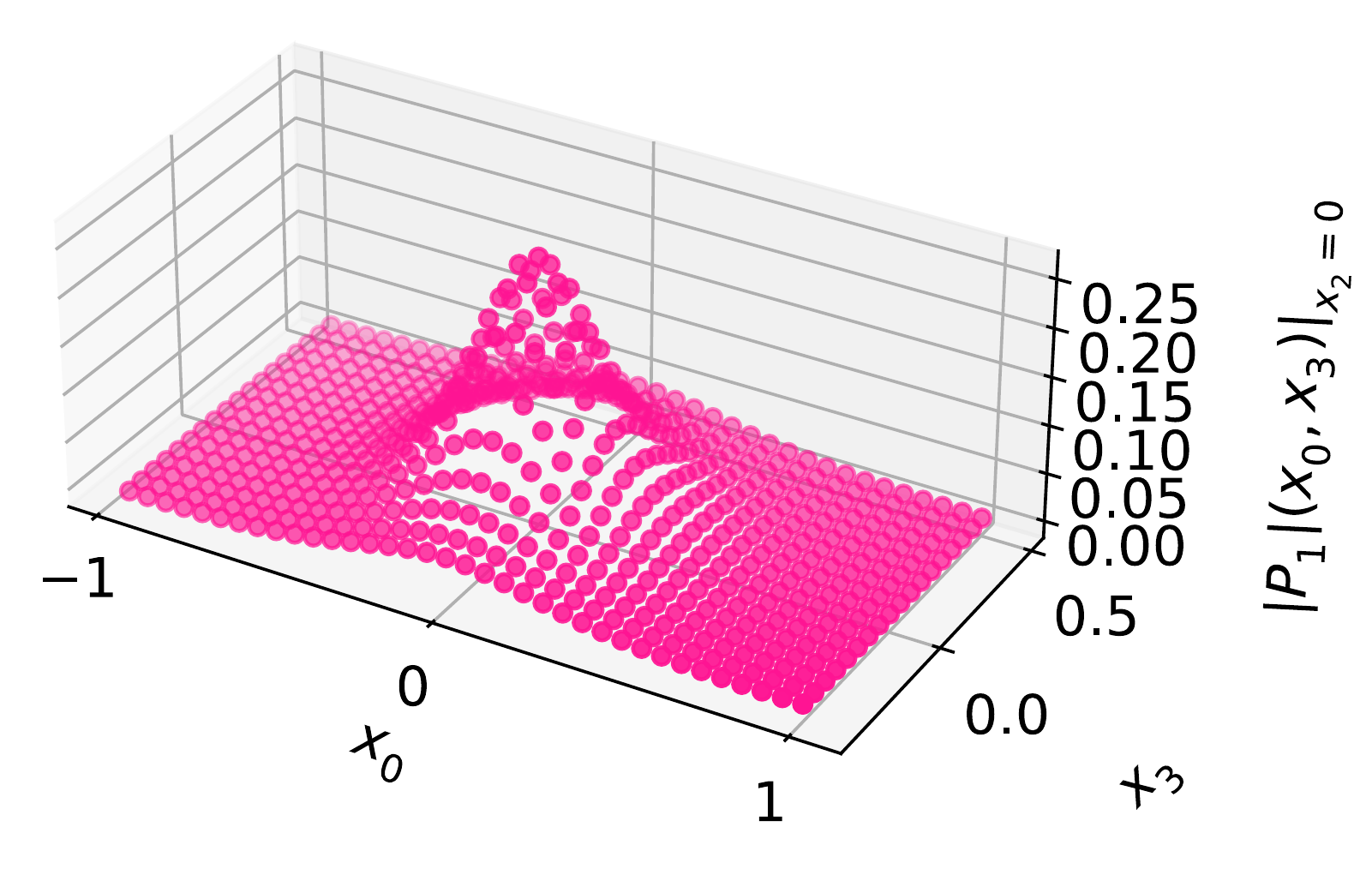}
  \caption{SU(3)}
  \label{fig:ploop1_13}
\end{subfigure}%
\begin{subfigure}{.5\textwidth}
  \centering
  \includegraphics[width=\linewidth]{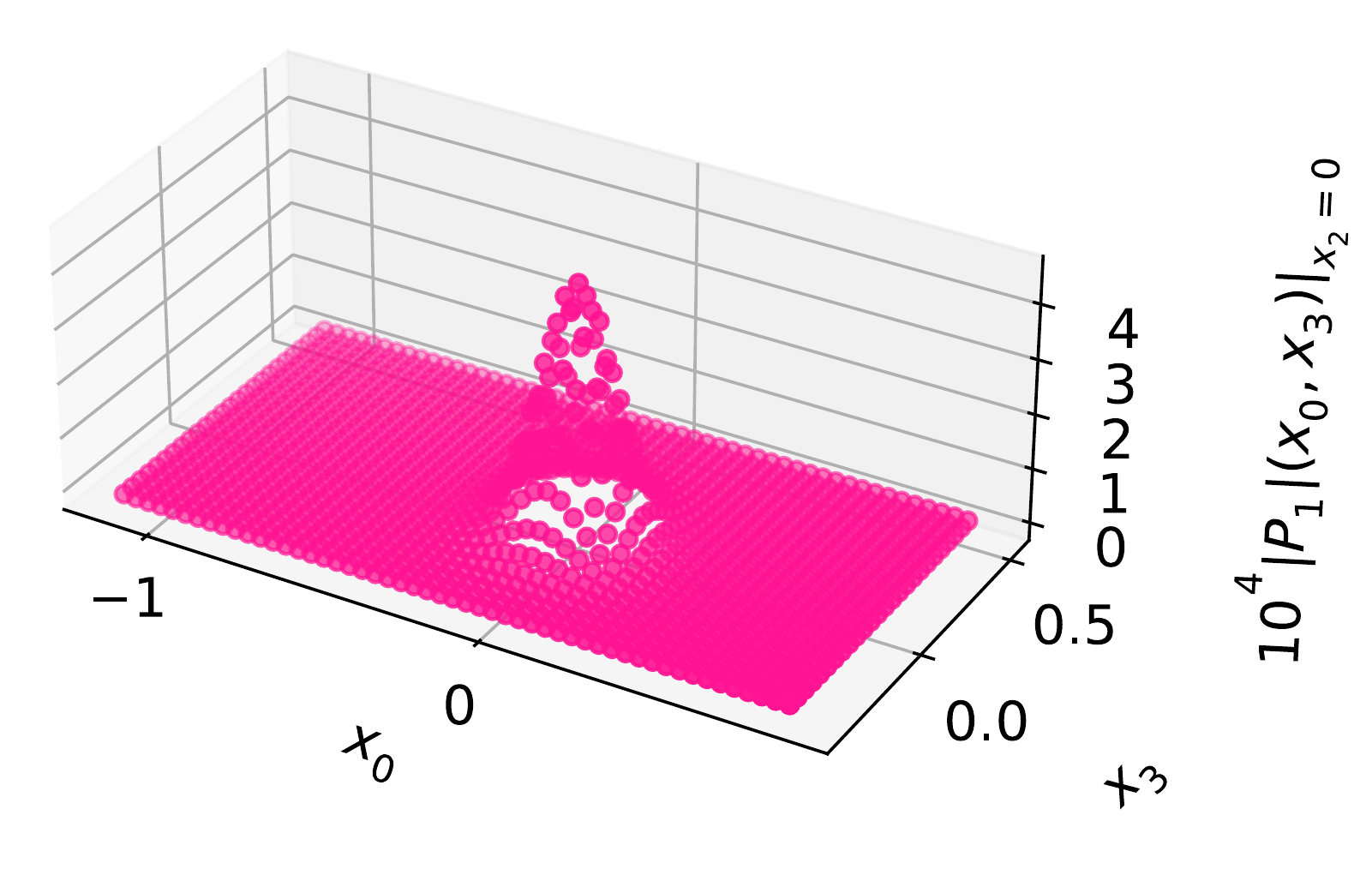}
  \caption{SU(13)}
  \label{fig:ploop1_3}
\end{subfigure}
\caption{ Modulus of $P_1$, the Polyakov loop winding once in the $x_1$ direction, as a function of the
coordinates in the 03 plane for $x_2$ at the instanton center. Data points have been shifted
so as to have the maximum of the action profile at $x=0$. Note the change of vertical scale in the case of $SU(13)$.}
\label{fig:ploop1}
\end{figure}

\begin{figure}[t]
  \centering
  \includegraphics[width=0.9\linewidth]{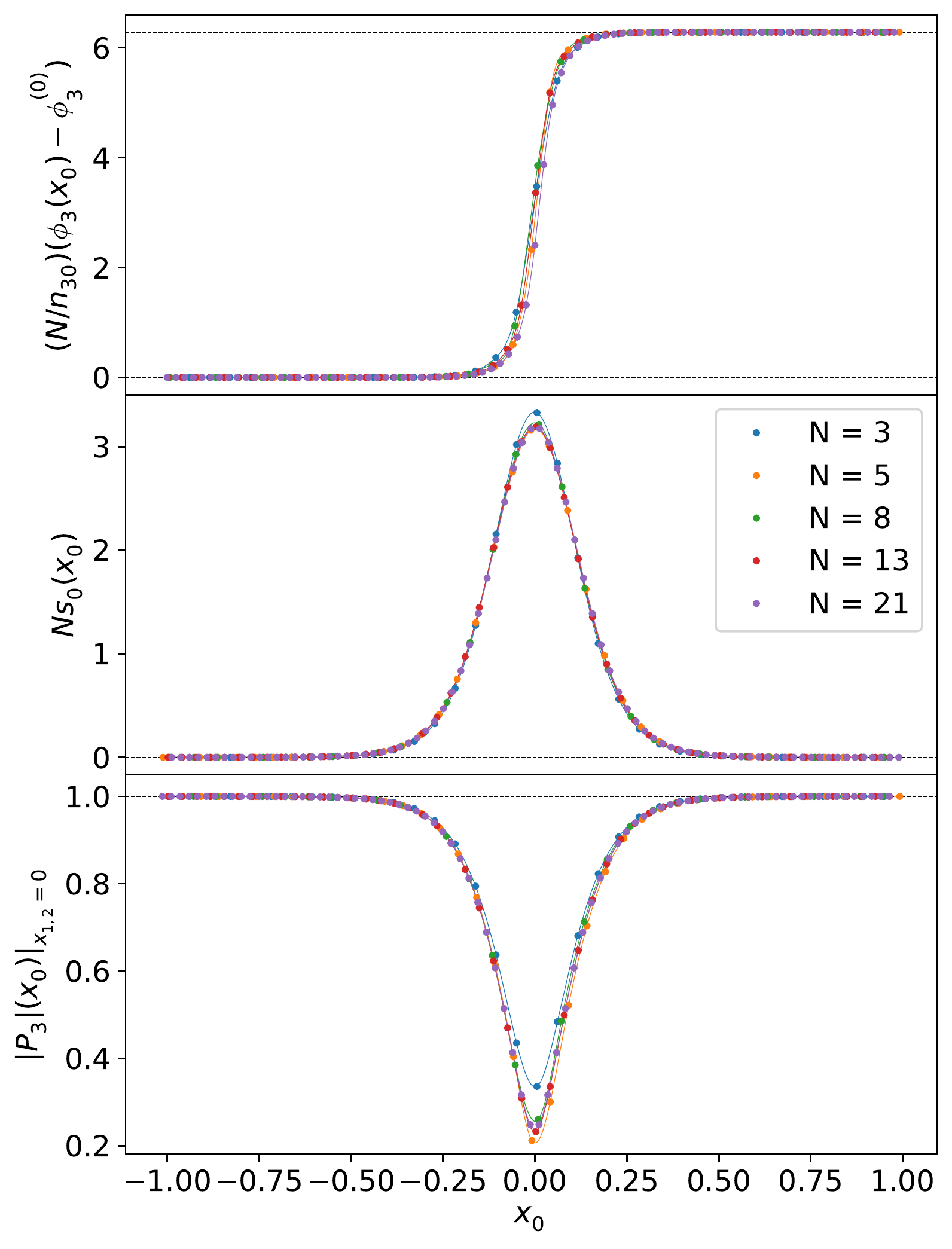}
\caption{For a torus of size $l_0=2l_3=2l$, we display as a function of time and from top to bottom: the phase of the Polyakov loop $P_3$ multiplied by $N/n_{30}$ (after subtracting the required powers of $2\pi/N$ so as to have the phase at $x_0=-l_0/2$ close to zero), the energy-profile of the instanton, and the modulus of $P_3$. The phase and modulus of the Polyakov loop are evaluated at: $x_1=x_2=0$.}
\label{fig:P3}
\end{figure}

Starting with the holonomies in the two short directions,  figure~\ref{fig:ploop1} displays, for gauge groups $SU(3)$ and $SU(13)$ and time extent $l_0=2 l$, the modulus of $P_1$ as a function of the coordinates in the $03$ plane for $x_2$ at the instanton center. Data points have been shifted so as to have the maximum of the action profile at $x=0$. The value of $|P_1|$ remains everywhere very small and tends to zero at $x_0 \rightarrow \pm \infty$, as expected. The same behaviour is observed for $P_2$ and all gauge groups.

We move next to describe the results for $P_3$. In fig.~\ref{fig:P3}  we display from top to bottom:
the phase of the Polyakov loop  multiplied by $N/n_{30}$, the energy-profile as a function of $x_0$, and the Polyakov loop modulus, all for torus sizes corresponding  $l_0=2l_3$. We have subtracted the required powers of $2\pi/N$ so as to have the phase at $x_0=-l_0/2$ close to zero in all cases. For the plot, Polyakov loops are computed with $x_1$ and $x_2$ taken as close as possible to the origin, we have observed however that the dependence on $x_1$ and $x_2$ is very small and the phase becomes independent of these coordinates in the large $N$ limit.  
The results converge in this limit to a universal function of $x_0$ which exhibits all the properties discussed previously, i.e.: 
\begin{itemize}
    \item 
Far from the instanton center the configuration approaches a vacuum solution with $P_1=P_2=0$ and $P_3$ equal to a root of unity, exemplified by the fact that the modulus tends to one and the phase becomes an integer multiple of $2 \pi /N$. 
\item
The fractional instanton interpolates between two vacuum configurations differing in phase by $2 \pi n_{30} /N$, as dictated by the boundary conditions. With our choice of twist, the jump in phase is given in absolute value by $ 2 \pi F_{n-2}/F_n$ and tends in the large $N$ limit to $2 \pi \varphi^{-2}$, with $\varphi$ the Golden Ratio.
\end{itemize}

\begin{figure}[t]
  \centering
  \includegraphics[width=0.9\linewidth]{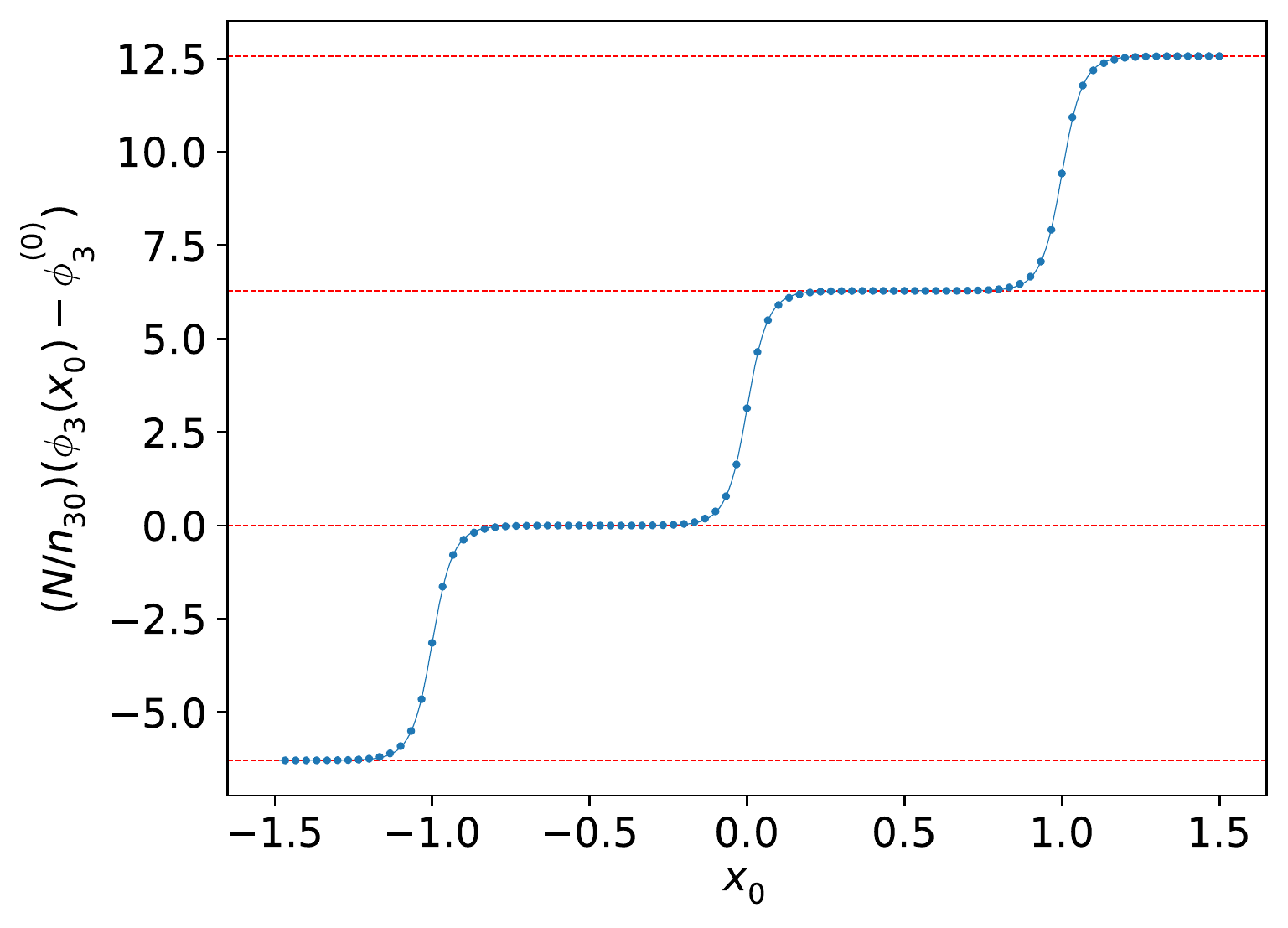}
\caption{We display, for the $SU(3)$ instanton configuration with $l_0=l_3$, the phase of the Polyakov loop $P_3$ multiplied by $N/n_{30}$ compared to the ansatz of eq.~\eqref{eq:phasePL} with $\xi= 0.999999999535$.}
\label{fig:phaseN}
\end{figure}

We have been able to find an ansatz that describes very well the time dependence of $\phi_3$. It is given by the following functional form:
\begin{align}
\phi_3(x_0) &= \phi_3^{(0)} +  \frac{n_{30}} {N} \,\left \{ \pi +2 \, {\rm am}\left(2 K(\xi) \frac{x_0-x_0^{(0)}}{l_0}, \xi \right)  \right \}
\label{eq:phasePL}
\end{align}
where am$(x,\xi)$ denotes Jacobi's Amplitude Function~\cite{DLMF} and 
\be
K(\xi) = \int_0^{\pi/2} \frac{d\theta}{\sqrt{1-\xi^2 \sin^2\theta}}.
\ee
with $\xi$ a tunable parameter taking values between 0 and 1.
Using the periodicity properties of the amplitude function:
\be
{\rm am}(x + 2 K(\xi), \xi)= {\rm am}(x , \xi) + \pi,
\ee
it is trivial to show that the phases transform as required under displacement by a torus period. An illustration of the quality of the fit for $N=3$, $l_0=l_3$ and $\xi= 0.999999999535$ is presented in fig.~\ref{fig:phaseN}.

\subsection{Wilson loops}

Finally, and continuing with the discussion of gauge invariant quantities, we proceed with the analysis of Wilson loops. We denote by $W_C(r)$ the Wilson loop defined as:
\be
W_C(r) = \frac{1}{N} \Tr \Big( P\exp \Big\{ -i \int_C dx_\mu A_\mu(x)\Big \}\, \Big ) \equiv |W_C(r) | \, e^{i\delta_C(r)},
\label{eq:wil}
\ee
where $C$ stands for a $T\times R$ square loop in the 03 plane, with $T=R=r$,  centered around the instanton locus at $x=0$. 

\begin{figure}[t]
\begin{subfigure}{.5\textwidth}
  \centering
  \includegraphics[width=\linewidth]{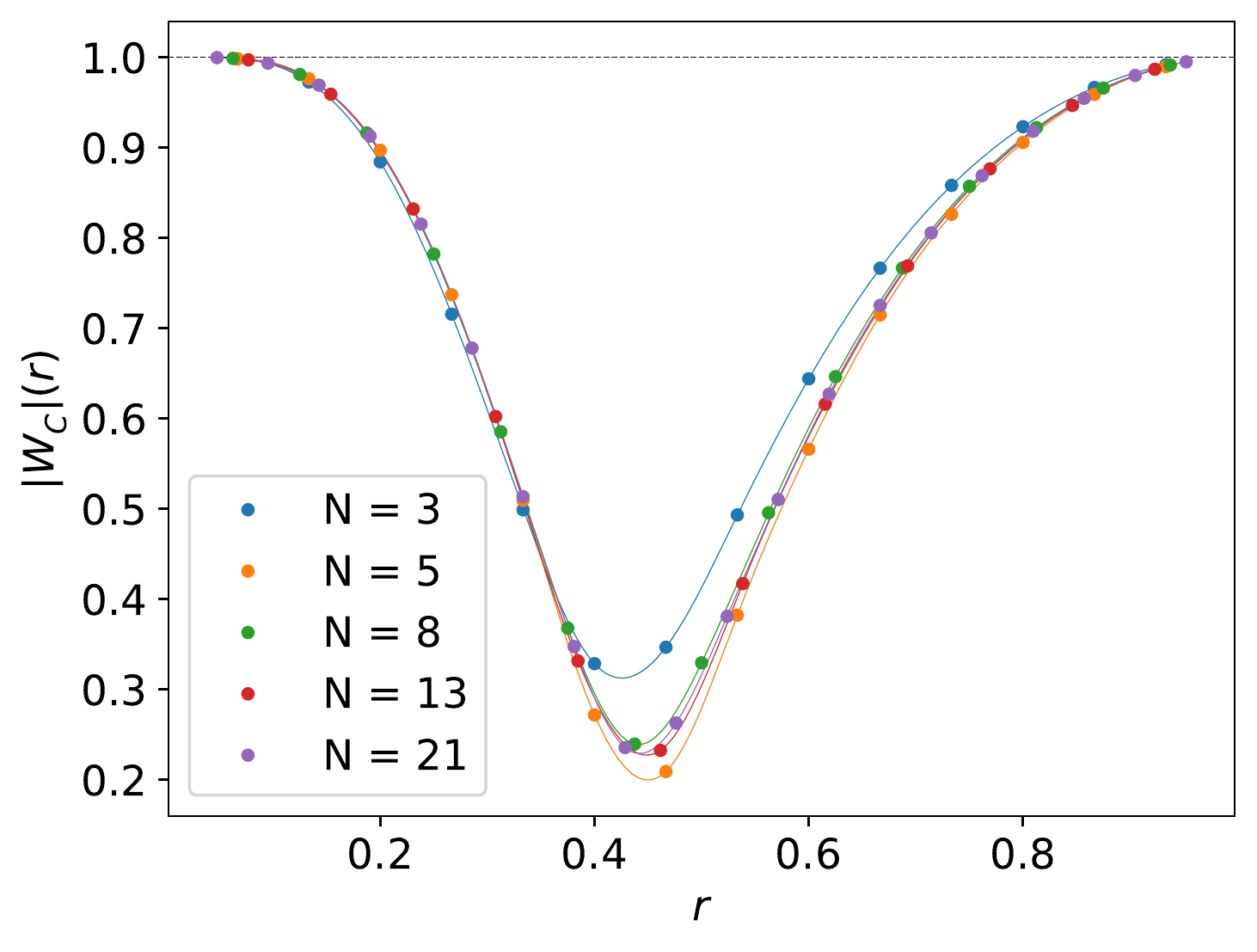}
\end{subfigure}%
\begin{subfigure}{.5\textwidth}
  \centering
  \includegraphics[width=\linewidth]{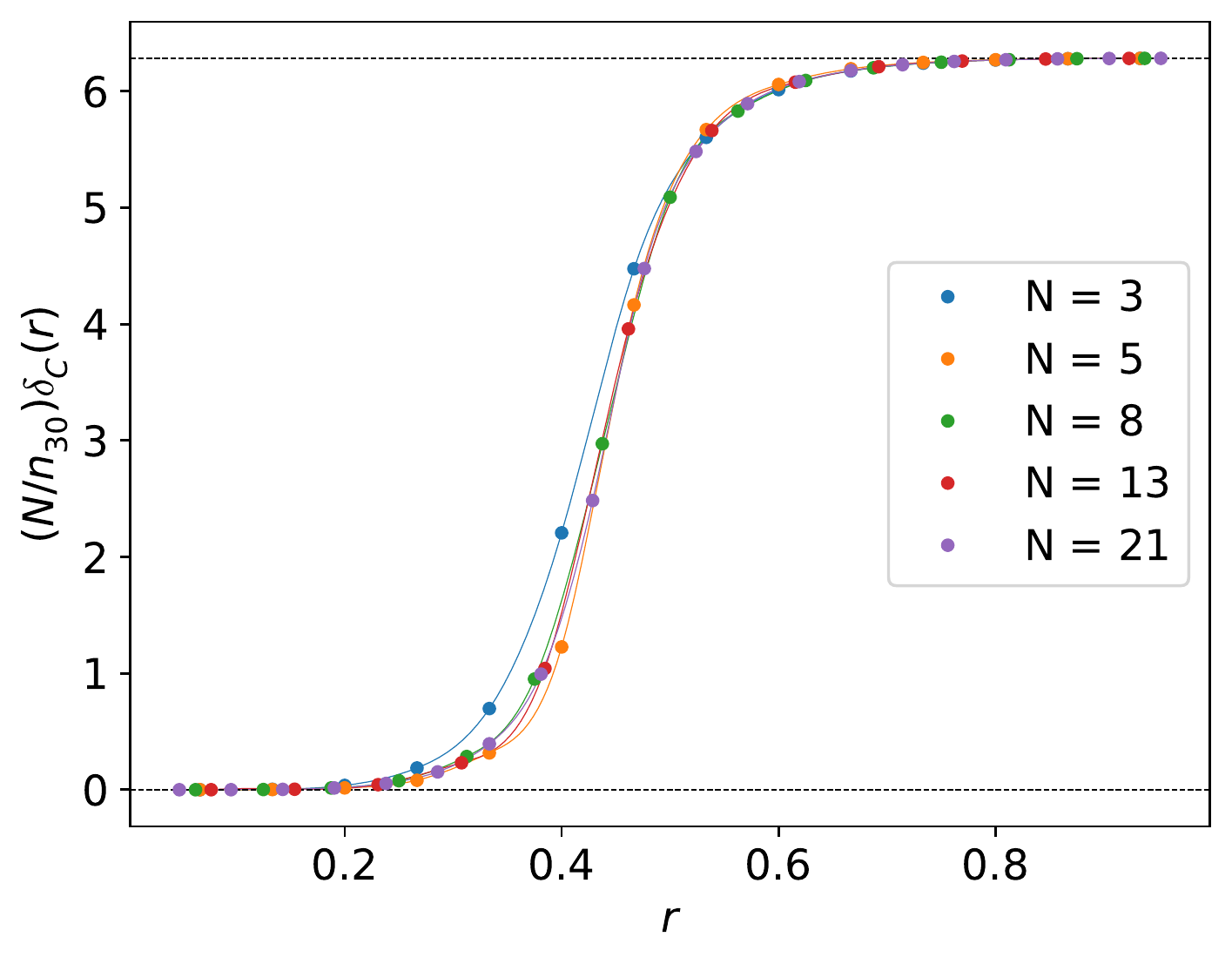}
\end{subfigure}
\caption{We display the modulus (left) and phase  (right), multiplied by $N/n_{30}$, of a square Wilson loop of side $r$ centered at the instanton position.  }
\label{fig:Wil-mod}
\end{figure}

Figure~\ref{fig:Wil-mod} displays as a function of $r$ the Wilson loop modulus and phase (normalized by a factor $N/n_{30}$)  for different gauge groups and $l_0=l_3=l$. Several remarks can be extracted from these plots. The modulus starts at one for $r=0$, and, as the distance $r$ grows, it decreases, reaches a minimum around $r\sim l/2$ and increases again back to 1, becoming at large $r$ a pure complex phase.  On the other hand, the phase jumps from $\delta_C(r)=0$ close to the instanton center to $\delta_C(r)=2\pi n_{30}/N$ at $r=l$. This shows that the Wilson loop around the fractional instanton encloses a non trivial $\mathbf{Z}_N$ flux -- notice that in the large $N$ limit the corresponding phase is dictated by the non-commutativity parameter $|\hat \theta|= F_{n-2}/F_n$ and approaches $\varphi^{-2}$ with $\varphi$ the Golden Ratio.

 \section{Conclusions}
    \label{s:conclusions}
    In this paper we have obtained numerical instanton-like solutions for gauge group $SU(N)$ with fractional topological charge $Q=1/N$. They have minimal action $S=8 \pi^2/N$ and are to a very large degree (anti)self-dual. They have been obtained on a 4-torus with twisted boundary conditions given by the magnetic flux $\vec m = (1,1,m)$ and $\vec k =(0,0,-\barm)$ where $N$ and $m=|\barm|$ are taken as the $nth$ and $nth-2$ integers in the Fibonacci sequence, i.e. $N=F_n$ and $m=F_{n-2}$. The advantage of this choice is that it is expected to prevent the appearance of tachyonic instabilities and $\mathbf{Z}_N\times \mathbf{Z}_N $ symmetry breaking  in the large $N$ limit.

We have analyzed the scaling of the solutions at large $N$. The aim was to find fractional charge solutions in the $\mathbf{R} \times \mathbf{T}^3$ Hamiltonian limit representing vacuum-to-vacuum tunneling events. Given that the twist induces an effective entanglement of colour and spatial degrees of freedom, we have selected different periods in the planes pierced by non-zero magnetic flux, taking $l_3=l$ and $l_1=l_2=l/N$ and analyzed the resulting solutions in the limit in which $l_0 \rightarrow \infty$. Considerations based on TEK reduction and continuum volume independence indicate that the effective size of the spatial torus obtained in this way is $l^3$. The resulting configurations scale in the large $N$ limit in agreement with this expectation. Action densities become independent of the twisted coordinates ($x_1, x_2$) and are localized in time with a width determined in terms of the effective length $l$.
The width of the energy profile is actually close to the maximal time-period for which our asymmetric torus supports abelian constant curvature self-dual solutions. 

The scaling of various other physical quantities in the large $N$ limit has been analyzed, including Polyakov and Wilson loop operators. They also scale as expected. The former show how the fractional instanton interpolates in time between two flat connections on the three torus, with a profile that, after an appropriate scaling, becomes $N$ independent in the large $N$ limit. The Wilson loop on the $03$ plane is non-trivial around the instanton, which acts effectively, at large distance, as a $\mathbf{Z}_N$ flux. In the large $N$ limit the value of the flux is given by the non-commutativity parameter $|\hat \theta|= |m|/N=F_{n-2}/F_n$ and approaches $\varphi^{-2}$ with $\varphi$ the Golden Ratio.

Our solutions share some properties with other $SU(N)$ configurations obtained previously in the literature. Being independent of two of the spatial coordinates and localized in the other two, they resemble the vortex-like structures on $\mathbf{T}^2 \times  \mathbf{R}^2$ obtained in refs.~\cite{Gonzalez-Arroyo:1998hjb,Montero:1999gq,Montero:2000pb}, however the latter correspond to an $N$-independent choice of twist $\vec k = \vec m = (1,0,0)$ and show a different large $N$-scaling of the energy profiles and Polyakov and Wilson loop operators. The dependence on $N$ of our energy profiles corresponds instead to the one observed for $SU(N)$ instantons on $\mathbf{T}^3 \times  \mathbf{R}$ for a symmetric spatial torus with twist $\vec m = (1,1,1)$, $\vec k =(1,0,0)$~\cite{GarciaPerez:1997fq,Montero:2000mv}. It is unclear to us at this point what determines the $N$ dependence in these other cases, and how it is related to the particular choice of twist,  a question that certainly deserves further investigation.

Let us finally point out that fractional instantons are relevant in the quest for analytical computability in Yang-Mills theories, a program that, as indicated in the introduction,  has received renewed attention in recent years. Including their contribution to the partition function is essential to extend analytical calculations on a twisted torus beyond the regime of perturbation theory  ($\Lambda l<<1$) -- see Refs.~\cite{RTN:1993ilw,GarciaPerez:1993jw,Gonzalez-Arroyo:1995ynx,Gonzalez-Arroyo:1995isl,Gonzalez-Arroyo:1996eos} where this program has been partially pursued for the case of $SU(2)$. Whether this allows to bridge the gap into the confinement regime remains so far an open question, a way to go may be to exploit the simplifications attained using volume independence and large $N$ dynamics as done in this work .

    \section*{Acknowledgments}
We thank Jos\'e L.F Barb\'on and Antonio Gonz\' alez-Arroyo for many useful discussions and comments to the manuscript. We are particularly grateful to Alberto Ramos for the $SU(N)$ lattice code for the flow  used in this work. This work is partially supported by grant PGC2018-094857-B-I00 funded by MCIN/AEI/10.13039/501100011033 and by “ERDF A way of making Europe”, and by the Spanish Research Agency (Agencia Estatal de Investigación) through grants IFT Centro de Excelencia Severo Ochoa  SEV-2016-0597 and  No CEX2020-001007-S, funded by MCIN/AEI/10.13039/501100011033. We also acknowledge support from the project H2020-MSCAITN-2018-813942 (EuroPLEx) and the EU Horizon 2020 research and innovation programme,
STRONG-2020 project, under grant agreement No 824093. 
JDG acknowledges support under grant PRE2018-084489 funded by MCIN/AEI/ 10.13039/501100011033 and, by “ESF Investing in your future”.
We acknowledge the use of the Hydra cluster at IFT and HPC resources at CESGA (Supercomputing Centre of Galicia).

\appendix
    \section{Numerical minimization of the action}
    \label{s:numerical}
    
In this Appendix we provide some details about the numerical procedure used to obtain the solutions described in section~\ref{s:solutions}. We also discuss some technical details about how to implement twisted boundary conditions on the lattice, and the type of lattice observables used for the calculation of the action density and the topological charge.

Lattice approximants of the fractional instanton configurations can be obtained minimizing the lattice equations of motion. 
There are various standard ways in the literature to attain this goal, 
in this work we have made use of the so-called gradient flow~\cite{Narayanan:2006rf,Lohmayer:2011si,Luscher:2009eq,Luscher:2010iy}. In the continuum, the flow is a smoothing procedure that
drives the gauge field towards a solution of the classical equations of motion by
 evolving the original gauge field $B_\mu(x,0)\equiv A_\mu(x)$ along the steepest descent through the flow equations:
\be
\partial_\tau B_\mu(x,\tau) = D_\nu G_{\nu \mu}(x,\tau),
\ee
where $D_\mu$ and $G_{\nu \mu}$ stand for the covariant derivative and field strength tensor of the flowed gauge field,
given by:
\begin{align}
D_\nu  G_{\nu \mu} &= \partial_\nu G_{\nu \mu} - i [ B_\nu, G_{\nu \mu}], \\
G_{\nu \mu} &= \partial_\nu B_\mu - \partial_\mu B_\nu - i [B_\nu, B_\mu].
\end{align}
On the lattice the flow equations can be easily discretized, in our study we have made used of the so called Wilson flow
discretization, which relies on the minimization of Wilson's plaquette action, to be described below. It is well known~\cite{GarciaPerez:1993lic} that for the Wilson action the leading ${\cal O} (a^2)$ lattice corrections to the continuum action are negative definite, therefore in the absence of a twist the only stable solution of the lattice equations of motion is the vacuum configuration. However, in the presence of a non-orthogonal twist, 
a pure gauge configuration is not compatible with the boundary conditions. In this case, the absolute minimum of the lattice action is different from zero and, for our choice of twist ($N=F_n$, $m=F_{n-2}$ and $k=-\barm=(-1)^{n+1} F_{n-2}$)  corresponds to a discretized version of the $Q=1/N$ fractional instanton, with an action equal, up to lattice artefacts, to $8\pi^2/N$.

In order to implement the concrete geometry we are interested in, we discretize the SU(N) gauge theory on a $ s NL \times L^2 \times NL $, corresponding to a continuum torus of size $sl \times (l/N)^2 \times l  $, with $l=NLa$ ($a$ the lattice spacing) and $s$ a positive integer.

To determine lattice observables as the action density, we have considered various various ways  of discretizing the continuum action constructed in terms of $1\times 1$ and $2\times 2$ plaquettes and given by~\cite{GarciaPerez:1993lic}:
\be
S(\epsilon)=\frac{4-\epsilon}{3}\sum_{n,\mu,\nu}\Tr\left(\mathbbm{I}- Z_{\mu \nu}^* (n) \, \plaq\right)
+\frac{\epsilon-1}{48}
\sum_{x,\mu,\nu}\Tr \left(\mathbbm{I}- \tilde Z_{\mu \nu}^* (n)\,  \twoplaq\right)\quad.
\label{eq:epsac}
\ee
In this construction the link variables are periodic and the twisted boundary conditions are imposed by introducing 
the twist-carrying center elements $Z_{\mu \nu}^*(n)$ and $\tilde Z_{\mu \nu}^* (n)$. The former is set to
one for $1\times 1$ all plaquettes except for the ones with coordinates $n_1 = n_2 = 0$, where we take:
\be
Z_{1 2} = Z_{2 1}^*= \exp \left  \{ \frac{i 2 \pi m}{N} \right \},
\ee
and for the ones with $n_0=n_3=0$ where we set instead:
\be
Z_{0 3} = Z_{3 0}^*= \exp \left  \{ \frac{-i 2 \pi \bar m}{N} \right \}.
\ee
As for $\tilde Z_{\mu \nu} (n)$, it is given by the product of all the single plaquette twist factors enclosed in 
the $2\times 2$ closed loop.
The factor $\epsilon$ controls the size of ${\cal O}(a^2)$ lattice corrections to the continuum action. 
Wilson's action corresponds to $\epsilon=1$, while the choice $\epsilon=0$ is taken so as to ensure that 
continuum and lattice classical actions  differ only by terms of ${\cal O}(a^4)$. 
The action density $s(x)$ used to display the profiles in section~\ref{s:solutions}, corresponds to the one for $\epsilon=0$, clover averaged over the four, $1 \times 1$ or $2 \times 2$, plaquettes attached to a given lattice point $n$ in each plane. 

The observable used to determine the topological charge is constructed out of a lattice discretization of the field strength tensor defined as::
\begin{align} 
F^{\rm cl}_{\mu\nu}(n) \nonumber &= -\frac{i}{8} \{
Z^*_{\mu\nu}(n) P_{\mu\nu}(n) + Z^*_{\mu\nu}(n-\hat{\nu})
P_{-\nu\mu}(n) \nonumber \\ & + Z^*_{\mu\nu}(n-\hat{\mu})
P_{\nu-\mu}(n) + Z^*_{\mu\nu}(n-\hat{\mu}-\hat{\nu})
P_{-\mu-\nu}(n) -c.c. \} ,
\label{eq.gmunu}
\end{align} with $P_{\mu\nu}(n)$ denoting the $1\times 1$ plaquette ($U_{-\mu}(n)\equiv U_\mu^\dagger(n-\hat \mu)$). In terms of $F^{\rm cl}_{\mu\nu}$,  the topological charge $Q$ is determined from: 
\be 
Q= \frac{1}{16 \pi^2} \sum_n \Tr \left \{
F^{\rm cl}_{\mu\nu}(n) \widetilde F^{\rm cl}_{\mu\nu}(n) \right \}
\label{eq:qlatt}
\ee 
This expression has lattice artefacts of ${\cal O}(a^2)$.

Finally, let us mention that one practical issue that influences the rate of convergence towards the absolute minimum of the action is the choice of initial configuration, $B_\mu(x,0)$. A random start is in most cases not practical since, particularly for large $N$, the flow tends to get easily trapped in metastable configurations with action higher than the minimum.   
More efficient is to start for instance from a twist eater configuration corresponding to an orthogonal twist with 
either $m$ or $\barm$ set to zero, slightly heated with a very large value of the lattice inverse 't Hooft 
coupling $b=1/\lambda$. In this way, we have been able to generate configurations of minimal action for 
the set of values of $N$ and lattice sizes indicated in table~\ref{tab:conf}.

    \FloatBarrier
\providecommand{\href}[2]{#2}\begingroup\raggedright\endgroup


\end{document}